\documentclass[preprint]{aastex}

\usepackage{natbib}
\usepackage{lineno}
\usepackage{lscape}
\usepackage{amssymb,amsmath}
\usepackage{soul}
\usepackage{chngcntr}

\shorttitle{Predictability of flux-transport solar dynamo models} 
\shortauthors{Sanchez et al.}

\newcommand{\erf}{{\mathrm{erf}}}

\begin{document}



\title{The predictability of advection-dominated flux-transport solar dynamo models}


\author{Sabrina Sanchez, Alexandre Fournier and Julien Aubert}
\affil{Institut de Physique du Globe de Paris, Sorbonne Paris Cit\'e, 
Universit\'e Paris Diderot\\ 
                UMR 7154 CNRS, F-75005 Paris, France}
\email{ssanchez@ipgp.fr}




\begin{abstract}
Space weather is a matter of practical importance in our modern society. Predictions of forecoming solar cycles mean amplitude and duration are currently being made based on flux-transport numerical models of the solar dynamo.  
Interested in the forecast horizon of such studies, we quantify the 
predictability window of a representative, advection-dominated, 
flux-transport dynamo model by investigating its sensitivity to initial 
conditions and control parameters through a perturbation analysis. We measure the 
rate associated with the exponential growth of an initial perturbation of the model trajectory, 
which yields a characteristic time scale known as the $e$-folding time $\tau_e$. The $e$-folding time is shown to decrease with the strength of the $\alpha$-effect, and to increase with the magnitude of the imposed meridional 
circulation. Comparing the $e$-folding time with the solar cycle periodicity, we obtain an average estimate for $\tau_e$  equal to $2.76$ solar cycle durations. From a practical point of view, the 
perturbations analysed in this work can be interpreted as uncertainties 
affecting either the observations or the physical model itself. 
  After reviewing these, we discuss their implications for solar cycle prediction.
%
%
 
\end{abstract}



\section{Introduction}

The Sun is a magnetic active star, which undergoes successive phases of high and low
magnetic activity with a  quasi-periodicity of approximately 11~years, powered by
a natural dynamo mechanism \citep{moffatt1978field}. This magnetic activity
 encompasses the recurrent manifestation of dynamical phenomena at the solar surface and in its
 atmosphere, such as sunspots, flares and coronal mass ejections \citep{priest1982solar}.
In addition to its remarkable regularity, solar activity exhibits longer term (decadal to centennial)
fluctuations \citep{hathaway2009solar}, and occasional periods
 of long-lasting near-quiescence, such as the Maunder Minimum.
 Since the solar cycle affects the energy radiated by the Sun, its understanding
 is key in elucidating the potential
control of solar activity on the long-term variability of the Earth's climate \citep{haigh2003effects}.

 Solar activity influences the terrestrial
 environment in other important aspects, connected with  the
 operation of satellites \citep{baker2000occurrence}, and the occurrence
 of geomagnetic storms, which can damage electric power grids and interfere with
radars and radio communications. These important issues highlight the strong need
for an accurate prediction of solar magnetic phenomena, which is one of the main goals
of space weather \citep{pulkkinen2007space}. Up until recently, such forecasting
 exercices were mostly conducted within an entirely data-driven framework,
 based for instance on geomagnetic precursors
methods \citep{hathaway2009solar,wang2009understanding}.
 It is sensible to believe, though, that more accurate and effective
 predictions could be obtained by combining these data with physical
 models of the Sun, using data assimilation \citep[e.g.][]{talagrand1997assimilation}.
 The most salient illustration of the application of data assimilation
  emanates every day
 from numerical weather prediction (NWP) centers, in the form of weather forecasts
 \citep[consult e.g.][for a historical perspective on NWP]{kalnay2003atmospheric}.
Application of data assimilation in geoscience also include oceanography
\citep[e.g.][]{brasseur06}, the study of
air quality \citep[e.g.][]{elbern2010inverse}, and land surfaces \citep[e.g.][]{houser2010land}.
In a context similar to that of the solar dynamo, data assimilation
 has  also recently come to the fore for the study of the Earth's dynamo, a surge
 motivated by our increased ability to observe and simulate the geomagnetic field
 \citep[e.g.][]{fournier2007case,fournier2010introduction,aubert2011npg,fournier2013ensemble}.
 Over the past 15 years, the study of the solar dynamo has witnessed an even more spectacular
  increase in its observational and modelling capabilities.
 The question of the feasibility of applying data assimilation techniques
 to the solar dynamo was asked a few years ago \citep{brun2007towards}, and was followed
 by a series of studies bearing promises \citep{kitiashvili2008application,
jouve2011assimilating,dikpati2012evaluating}.

\begin{figure}[!ht]
\epsscale{0.85}
\plotone{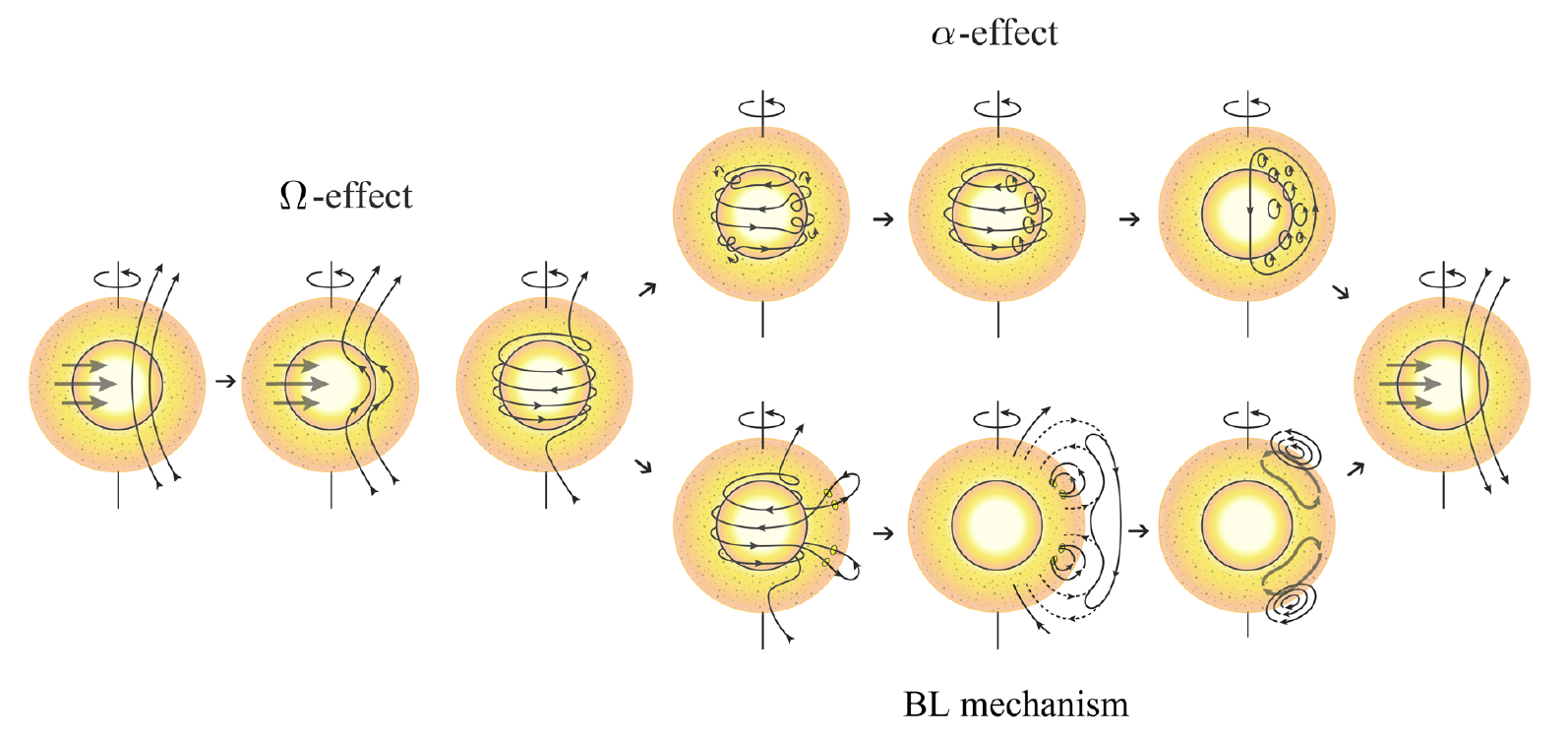}
\caption{
Sketch of the main processes at work in our solar dynamo model.
The $\Omega$-effect (left) depicts the transformation of a primary poloidal field
into a toroidal field by means of the differential rotation. The poloidal field regeneration
is next accomplished either by the $\alpha$-effect (top) and/or by the Babcock-Leighton mechanism (bottom).
In the $\alpha$-effect case, the toroidal field at the base of the convection zone is
subject to cyclonic turbulence.  Secondary small-scale poloidal 
fields are thereby created,
and produce on average a new, large-scale, poloidal field. 
 In the
Babcock-Leighton mechanism, the primary process for poloidal field regeneration
is the formation of sunspots at the solar surface from the rise of buoyant toroidal magnetic flux tubes from
the base of the convection zone. The magnetic fields of those sunspots nearest to the equator 
in each hemisphere diffuse and reconnect, while the field due to those 
sunspots closer to the poles
has a polarity opposite to the current one, which initiates a polarity reversal. 
The newly formed polar magnetic flux is transported by the meridional flow to the 
deeper layers of the convection zone, thereby creating a new large-scale poloidal field.}
\label{fig_dynamo}
\end{figure}

The physical model of the solar dynamo that should enter this inverse problem
 machinery remains to be defined. Forward modelling of the solar dynamo
 has shed light on the main physical processes believed to be
responsible for the solar cycle \citep[][for a review]{charbonneau2010dynamo}.
 Kinematic dynamo theory stresses that these processes are connected with
 the continuous transformation of poloidal magnetic energy into toroidal
 magnetic energy (the $P\rightarrow T$ conversion), and vice-versa (the
 $T\rightarrow P$ conversion, necessary to close the dynamo loop).
 There is now little doubt that the $\Omega$-effect, which denotes the shearing action of
the differential rotation of the plasma flow, is responsible for
 the $P\rightarrow T$ conversion.
 Through the advent of helioseismology, the large-scale, interior,
 differential rotation was mapped in detail \citep{tomczyk1995measurement},
 which made it possible to infer that the most likely location of the $\Omega$-effect
 is the base the convection zone, a region known as the tachocline
\citep{howe2000dynamic}.
 There is less consensus regarding the processes at work behind the
 $T\rightarrow P$ conversion. The
   mean-field $\alpha$-effect \citep{parker1955hydromagnetic},
 and the Babcock-Leighton mechanism
 \citep{babcock1961topology,leighton1969magneto} are two commonly envisioned
 possibilities. The former rests on the
 large scale effect of small scale turbulent motions
 whose twisting action
 can transform a toroidal field line into a poloidal field line.
The latter relies on
empirical evidences of the process of diffusion and reconfiguration of the
 magnetic field of sunspots.
The three aforementioned processes ($\alpha$-effect, $\Omega$-effect, and
Bacbcock-Leighton mechanism) are illustrated in Figure~\ref{fig_dynamo}.

The ambiguity between the $\alpha$-effect scenario and
 the Babcock-Leighton mechanism would disappear, should one be in a position
 to carry out the full three-dimensional numerical integration
 of those equations governing the solar dynamo.
 Despite the monotonic and dramatic increase in compute power which
 already led to substantial achievements \citep[e.g.][]{brun2004global,charbonneau2013modeling},
 such a comprehensive integration remains out of reach due to the wide range of temporal and spatial scales
 induced by the
high level of turbulence expected inside the solar convection zone.
 On the other hand, and from a more practical perspective,
a large body of work has shown that axisymmetric mean-field solar dynamo models were
 able to reproduce
many of the observed features of solar activity \citep{charbonneau2010dynamo}. The most
recent and representative illustrations of this strand rely on the
advection of  magnetic flux by a
meridional flow (following in general the Babcock-Leighton mechanism). These models, called 
'flux-transport' models, 
are in particular successful in accounting for the equatorward migration of the solar toroidal field
and the observed phase-locking of the solar cycle
\citep{dikpati1999babcock,charbonneau2000stochastic}.

Such flux-transport models may make it possible to predict the amplitude
and duration of the upcoming solar cycles. The first
studies addressing this possibility
\citep{dikpati2006predicting,choudhuri2007predicting}
considered direct incorporation of data into models,
 essentially by imposing (in a strong sense) surface boundary values inherited from the data
 onto the model, whereas an assimilation scheme would require this to
 happen in a weak sense, through some flavour of the so-called best linear unbiased estimator
 (BLUE), whose goal is to combine in an optimal fashion the data and the model,
 considering the uncertainties affecting both.
 Independently of the data assimilation scheme one may resort to, and
 as good as it may be, there exists an intrinsic limit to its predictive power.
\citet{bushby2007predicting} point out that this limit arises either from the
stochastic nature of the Babcock-Leighton and $\alpha$-effects, or from nonlinear deterministic processes.
 They stress, in addition, that the lack of constraints on the 
  exact nature of the key physical mechanisms which sustain these models 
 and govern their time-dependency, such as the $\alpha$-effect,
 make their ability to capture the essentials of the solar dynamo process questionable. 
 They conclude that under the best circumstances of a near-perfect model, 
 the shape of the solar cycle could only
 be predicted one or two cycles ahead. As this best case scenario is out of reach, 
 they argue that a reliable forecasting exercise is untractable.

The same critic was made regarding weather prediction during its early years.
 The seminal work by \citet{lorenz1963deterministic} showed the extreme sensitivity
 of a deterministic system governed by a simple
 set of nonlinear coupled differential equations to its initial conditions.
In a subsequent study, \citet{lorenz1965study} estimated the time scale
 of
 divergence $\tau$ of two initially very close dynamical trajectories (termed
 {\sl twin} trajectories in the following) to be of a few days
 (Lorenz's simple model aimed at representing atmospheric convection).
 More realistic models of the atmosphere have now established that
 $\tau$ is equal to 2~weeks. This value has to be confronted with
 the current forecast horizon of NWP, which is (depending
 on the center)  between 7 and 9 days. The combined progress of observation,
 models, and data assimilation algorithms over the past 30 years
 has resulted roughly in a gain of one day per decade, bringing
 the operational limit closer and closer to the theoretical limit.
 
 One may wonder to which extent the progress made by the atmospheric
 community could be expected within the solar community. 
 Doing so, one immediately realizes that these two dynamical
  systems (the atmosphere and the Sun) are dramatically 
   different. Whereas the Earth's atmosphere is a thin and 
   directly observable layer, the solar convection zone 
   is an almost entirely concealed thick shell. 
   Moreover, the physics of the atmosphere is much better 
   constrained than that at work behind the solar dynamo 
   (consult \cite{vallis2006atmospheric} for a review of atmospheric 
   processes). Bearing these substantial differences in mind, and assuming 
   that the basic physics involved in the solar dynamo is faithfully 
   captured by mean-field models, one may still hope that the short-term 
   prediction of at least  some of the features of the 
   solar cycle (e.g. duration and mean amplitude) is  possible.
 
Knowledge of the modulations and mean intensity
of the upcoming solar cycles from mean-field models may serve as an important
 input for more specific space weather considerations. In this study,
 we therefore wish to adopt an operational perspective. Assuming that mean-field models
 will be effectively used to forecast solar activity, our goal
 here is to quantify their intrinsic limit of predictability $\tau$ (the equivalent
 of the 2~weeks discussed above for the atmosphere), following
 the methodology proposed recently by \cite{hulot2010earth} and \cite{lhuillier2011earth}
 in order to estimate $\tau$ for the Earth's dynamo.

This paper is organized as follows.
In section~\ref{section_model}, we describe our working mean-field model
 and detail its numerical implementation.
 We next inspect the sensitivity of this model to its control parameters
 in section \ref{section_sensitivity}.
 Section~\ref{section_eg} presents the systematic study of
  the error growth between twin trajectories. This allows us 
  to  evaluate $\tau$,
 and to assess its sensitivity to its control parameters.
 Finally, we discuss in section \ref{section_discussion} the influence 
 of modelling and observational errors on the
 practical limit of predictability of the model.


\section{The model and its numerical implementation} \label{section_model}

	Our flux-transport model is the one presented by \citeauthor{sanchezrevision} (in press); it includes both
     the $\alpha$ and Babcock-Leighton (BL) scenarios for the $T\rightarrow P$ conversion.
      The first reason for adding an $\alpha$-effect to a standard Babcock-Leighton 
      flux-transport model is that a dynamo running on a BL mechanism alone can not recover
        from a quiescent phase devoid of sunspots. As reported by 
         \citeauthor{sanchezrevision} (in press), the model set-up enables
          the appearance of a long-term variability (succession of active and quiet phases), 
           which can then be interpreted
    as the result of the competition between 
    the $\alpha$-effect operating at the tachocline 
    and a BL mechanism 
    operating at the solar surface.
      In addition, a deep location of the $\alpha$-effect is known to favor the sought 
        antisymmetrical evolution of the magnetic field
       in the Northern and Southern hemispheres \citep{dikpati2001flux,bonanno2002parity}.

\begin{figure}[h]
\epsscale{0.7}
\plotone{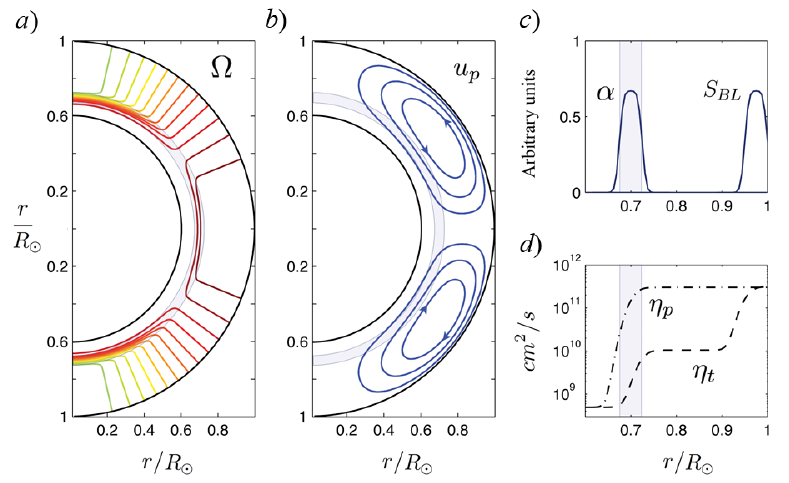}
\caption{ Components defining the class of solar
dynamo models used in this study: \textit{a}) Isocontours of the angular velocity $\Omega$;  
\textit{b}) Meridional circulation streamlines; \textit{c}) radial
profiles of 
the $\alpha$-effect and Babcock-Leighton poloidal source terms; 
\textit{d}) radial profiles of the magnetic diffusivities. In each panel, the shaded
regions symbolize the tachocline. Aside from the differential rotation, whose amplitude
remains fixed in this study, 
we vary the magnitude of these various components, whose relative contributions
are described by a suite of non-dimensional numbers (see text for details).}
\label{fig_flow}
\end{figure}

     Let us now write accordingly the modified mean-field induction equation \citep{moffatt1978field} for 
the large-scale magnetic field $\mathbf{B}$ 
\begin{eqnarray} \label{induction_eq}
\frac{\partial\mathbf{B}}{\partial t}=\boldsymbol{\nabla}\times \left[ \bold{U}\times\bold{B}
	-\eta\boldsymbol{\nabla}\times\mathbf{B}+\alpha\mathbf{B}
	 + S_{BL}B_{\varphi}
     \mathbf{{\hat{e}}_{\varphi}}
   \right],
\end{eqnarray}
      where ${\mathbf U}$ is the prescribed flow, 
      $\eta$ is the turbulent diffusivity,  
      $\alpha$ is the turbulent magnetic helicity, and 
       $S_{BL}B_{\varphi}\mathbf{{\hat{e}}_{\varphi}}$
        is the BL source term ($\mathbf{{\hat{e}}_{\varphi}}$ is the unit vector
     in the direction of longitude). We will specify the profiles of these
         various physical fields in the following. The definitions that we will need are summarized in 
        Table~\ref{table_symb} and the profiles shown in Figure~\ref{fig_flow}. 
   
\begin{table}[h]
\scriptsize
\begin{center}
\caption{Summary of the mathematical symbols used in the model, their values and a brief explanation of their meaning.}
\begin{tabular}{lrl}
\tableline
\tableline 
Symbol & Value & Interpretation \\
\tableline

$r_{tc}$ & $0.7 R_\sun$ &  Radial location of the center of the tachocline \\ 
$\delta r$ & $0.05 R_\sun$ & Thickness of the tachocline \\
$\Omega_{eq}$& $ 2\pi\times 460.7 $ nHz  & Rotation rate at the equator \\
$\alpha_0$   &  $0.34 - 1.03 $ m/s    & Strength of the $\alpha$-effect \\
$S_{BL_0}$       &  $0.02 - 0.06 $ m/s    & Strength of the Babcock-Leighton mechanism \\
$u_0$        & $13.27 - 17.68$ m/s & Velocity of the superficial meridional flow at mid-latitude\\
$\eta_r$     & $5\times10^8$ cm$^2$/s    & Effective diffusivity near the radiative zone \\
$\eta_{cz}$  & $1\times10^{10}$ cm$^2$/s & Effective diffusivity at the bottom of the convection zone \\
$\eta_s$     & $3\times10^{11}$ cm$^2$/s & Effective diffusivity at the solar surface \\

\tableline
\end{tabular}
\label{table_symb}

\end{center}
\end{table}

Under the assumption of axisymmetry, the magnetic and flow fields are further expressed in terms
of their poloidal and toroidal components in spherical coordinates $(r,\theta,\varphi)$ as
\begin{eqnarray}
 \bold{B}(\mathbf{r},t) &=& \boldsymbol{\nabla} \times  \left[ A_{\varphi}(\mathbf{r},t) \mathbf{\hat{e}_{\varphi}}  \right] 
	+ B_{\varphi}(\mathbf{r},t) \mathbf{{\hat{e}}_{\varphi}}, \label{poltor_B_ax} \\
 \bold{U}(\mathbf{r}) &=& \mathbf{u}_p (\mathbf{r}) + r \sin\theta  \, \Omega(\mathbf{r}) \mathbf{{\hat{e}}_{\varphi}},
\label{poltor_u}
\end{eqnarray}
in which $A_{\varphi}$ is the poloidal potential and $B_{\varphi}$ is the toroidal field. The prescribed
time-independent flow is defined by the angular velocity $\Omega$ and the meridional 
circulation $\mathbf{u}_p$, shown in Figures \ref{fig_flow}(\textit{a}) and \ref{fig_flow}(\textit{b}), respectively. 
Helioseismic data provide strong constraints on $\Omega$, which will thus remain fixed in 
the remainder of this work, and approximated using the analytic formula of \cite{dikpati1999babcock}.
On the contrary, the large-scale meridional circulation $\mathbf{u}_p$ remains poorly constrained. 
 For the sake of simplicity, we will follow the one-cell per hemisphere description of  \citet{dikpati1999babcock}.

The poloidal--toroidal decomposition of the magnetic and flow fields prompts us to define
poloidal and toroidal components for
 the turbulent diffusivity $\eta$, denoted by $\eta_p$ and $\eta_t$, respectively. This
  distinction rests on the analysis made by \cite{chatterjee2004full}, who pointed out
   that the toroidal field strength is expected to be much larger than the poloidal
    field strength troughout the convection zone. 
   This should decrease notably the efficiency of toroidal turbulent diffusion compared with
     its poloidal counterpart. With this distinction at hand, 
 injection of Eqs. (\ref{poltor_B_ax}) and (\ref{poltor_u}) into Eq. 
(\ref{induction_eq}) gives rise to a set of two coupled partial differential equations 
for $A_{\varphi}$ and $B_{\varphi}$
\begin{eqnarray}
\frac{\partial A_{\varphi}}{\partial t} + \frac{\mathbf{u}_p}{r \sin\theta}  \cdot \boldsymbol{\nabla} \left( r \sin\theta A_{\varphi} \right) &=&  {\eta}_p \left( \nabla^2 - \frac{1}{r^2 \sin^2\theta} \right) A_{\varphi} 
\nonumber \\ 
&+& \alpha (r,\theta; B_{\varphi}) B_{\varphi} +  S_{BL}(r,\theta;B_{\varphi}^{tc}) B_{\varphi}^{tc},
\label{induction_pol_dim} \\
\nonumber \\
\frac{\partial B_{\varphi}}{\partial t} + \; r\sin\theta \boldsymbol{\nabla} \cdot \left( \frac{\mathbf{u}_p 
	B_{\varphi}}{r \sin\theta}  \right) &=& {\eta}_t \left( \nabla^2 - \frac{1}{r^2 \; \sin^2 \theta} \right) B_{\varphi} 
    \nonumber\\
&+& \frac{1}{r} \frac{\partial {\eta}_t}{\partial r} \frac{\partial (r B_{\varphi})}{\partial r} +  r \; 
	\sin\theta \left( \boldsymbol{\nabla} \times A_{\varphi}\mathbf{\hat{e}_{\varphi}} \right) \cdot \left( \boldsymbol{\nabla} \Omega \right),
    \label{induction_tor_dim}
\end{eqnarray}
where $B_{\varphi}^{tc}=B_\varphi (r=r_{tc},\theta,t)$ is the toroidal field at the tachocline, 
 defined in this work as the spherical shell of mean radius $r_{tc}=0.7 R_\sun$, with a thickness
 $\delta r=0.05 R_\sun$. 
 The dependency of the $S_{BL}$ term in Eq.~(\ref{induction_pol_dim}) expresses 
the non-local character of the BL source term. 
Even if it is active within the surface layers, the BL regeneration process
 is thought to originate from processes occurring in the vicinity of the tachocline -- 
 numerical models indeed indicate that the formation of tilted bipolar regions at the surface 
  is mostly controlled by the strength 
of toroidal flux tubes prior to their buoyant instability
\citep{dsilva1993theoretical}. Their finite rise time should induce a time lag between
the onset of the instability and the formation of the bipolar regions, on the order
of some days to a few weeks \citep{jouve2010buoyancy}. We will neglect this
delay on the account of it being small compared to the time scales of interest here. 

Turning now our attention to the $\alpha$-effect,
we use the standard formula of $\alpha$-quenching, which is written as
\begin{eqnarray} \label{alpha_quench}
\alpha(\mathbf{r};B_{\varphi})=\frac{\bf \alpha_0}{1+{\left(\frac{B_{\varphi}}{B_{eq}}\right)}^2} f_{\alpha}(\mathbf{r}),
\end{eqnarray}
in which  $\alpha_0$ is a typical magnitude, $B_{eq}=10^4$~G \citep{fan2009magnetic} and $f_{\alpha}(\mathbf{r})$ restricts
 the $\alpha$-effect to the mid-latitudes of the tachocline, according to 
\begin{eqnarray} \label{alpha_prof}
f_{\alpha}(\mathbf{r})=\frac{1}{4}\left[1+\erf\left(\frac{r-r_1}{d_1}\right)\right]
	\left[1-\erf\left(\frac{r-r_2}{d_2}\right)\right]\cos\theta\sin\theta,
\end{eqnarray}
where $r_1=r_{tc}-\delta r/2$, $r_2=r_{tc} + \delta r/2$, and $d_1=d_2=0.01 R_{\sun}$. The radial 
variations of $f_{\alpha}$ are shown in Figure \ref{fig_flow}(\textit{c}).

The Babcock-Leighton $S_{BL}$ source term operates within bounds of the magnetic field strength
\citep{dsilva1993theoretical}, specifically between
$ B_{\varphi,\mathrm{min}}^{tc}~=~10^4$~G and $B_{\varphi,\mathrm{max}}^{tc}~=~10^5$~G. 
Denoting the magnitude of this source term by ${\mathbf S_{BL_0}}$, 
we write accordingly 
\begin{eqnarray} \label{BL_term}
S_{BL}(\mathbf{r};B_{\varphi}^{tc}) = \frac{{\mathbf S_{BL_0}}}{4} \left[ 1 + \erf \left( B_{\varphi}^{tc \; 2} - B_{\varphi, \mathrm{min}}^{tc \; 2} \right) \right] 
	\left[ 1 - \erf \left( B_{\varphi}^{tc \; 2} -  B_{\varphi, \mathrm{max}}^{tc \; 2} \right) \right] f_{BL}(\mathbf{r}).
\end{eqnarray}
The radial and latitudinal distribution $f_{BL}(\mathbf{r})$ is in turn
given by
\begin{eqnarray} \label{BL_profile}
f_{BL}(\mathbf{r})=\frac{1}{4}\left[1+\erf\left(\frac{r-r_3}{d_3}\right)\right]
        \left[1-\erf\left(\frac{r-r_4}{d_4}\right)\right]\cos\theta\sin\theta ,
\end{eqnarray}
where $r_3=0.95R_{\sun}$, $r_4=R_{\sun}$ and $d_3=d_4=0.01R_{\sun}$. 
The radial distribution of $f_{BL}$  is shown in Figure~\ref{fig_flow}(\textit{c}).

The poloidal and toroidal diffusivities in Equations (\ref{induction_pol_dim}) and (\ref{induction_tor_dim}), are written as
\begin{eqnarray}
\eta_p(r)&=&\eta_r +\eta_s \frac{1}{2}\left[1+\erf\left(\frac{r-r_5}{d_5}\right)\right],\label{diffP} \\
\nonumber \\
\eta_t(r)&=&\eta_r +\eta_{cz} \frac{1}{2}\left[1+\erf\left(\frac{r-r_6}{d_6}\right)\right]
                +\eta_s \frac{1}{2}\left[1+\erf\left(\frac{r-r_7}{d_7}\right)\right], \label{diffT} 
\end{eqnarray}
in which $r_5=0.7 R_{\sun}$, $r_6=0.72 R_{\sun}$, $r_7=0.95 R_{\sun}$, $d_5=d_6=d_7=0.025 
R_{\sun}$,  $\eta_r$ is the diffusivity at the boundary with the radiative zone, $\eta_{cz}$ is 
the diffusivity in the turbulent convection zone,
and $\eta_s$ is the diffusivity in the surface layers (which applies to the poloidal field
 over the entire convection zone). The radial profiles of the diffusivities
are shown in Figure~\ref{fig_flow}(\textit{d}). 
This model pertains to the generic class of advection-dominated models: owing
 to the low values of the diffusivities throughout the convection zone, the coupling
 between the regions where the poloidal and toroidal fields are generated is ensured
 by the meridional circulation. In diffusion-dominated 
 models \citep[e.g.][]{chatterjee2004full}, this
 coupling is on the contrary accomplished by turbulent diffusion.  
 
\par

In order to express the dynamo equations in their nondimensional form, we choose the solar radius $R_{\sun}$ as the
length scale and the magnetic diffusion time $R_{\sun}^2/\eta_s$ as the time scale (roughly equal to $500$~years).  This yields
\begin{eqnarray}
\frac{\partial A_{\varphi}}{\partial t} + \frac{\mathrm{Rm}}{r \sin\theta} \tilde{\mathbf{u}}_p \cdot \boldsymbol{\nabla} \left( r \sin\theta A_{\varphi} \right) &=&  \tilde{\eta}_p \left( \nabla^2 - \frac{1}{r^2 \sin^2\theta} \right) A_{\varphi} \nonumber \\ 
&+&\mathrm{C}_{\alpha}\tilde{\alpha}\,B_{\varphi}+\mathrm{C}_{BL} \tilde{S}_{BL}\,B_{\varphi}^{tc},
\label{induction_pol} \\
\nonumber \\
\frac{\partial B_{\varphi}}{\partial t} + \mathrm{Rm} \; r\sin\theta \boldsymbol{\nabla} \cdot \left( \frac{\tilde{\mathbf{u}}_p 
	B_{\varphi}}{r \sin\theta}  \right) &=& \tilde{\eta}_t \left( \nabla^2 - \frac{1}{r^2 \; \sin^2 \theta} \right) B_{\varphi} \nonumber\\
&+& \frac{1}{r} \frac{\partial \tilde{\eta}_t}{\partial r} \frac{\partial (r B_{\varphi})}{\partial r}+\mathrm{C}_{\Omega} \; r \; 
	\sin\theta \left( \boldsymbol{\nabla} \times A_{\varphi}\mathbf{\hat{e}_{\varphi}} \right) \cdot \left( \boldsymbol{\nabla} \tilde{\Omega} \right).
    \label{induction_tor}
\end{eqnarray}

Equations (\ref{induction_pol}) and (\ref{induction_tor}) contain six nondimensional numbers 
characterizing the relative importance of each term in the equations
\begin{eqnarray}
\mathrm{Rm}&=&u_oR_\sun/\eta_s, \label{eq:Rm} \\
\mathrm{C}_{\Omega}&=&\Omega_{eq}R_\sun^2/\eta_{s}, \label{eq:comega} \\
\mathrm{C}_{\alpha}&=&\alpha_0R_\sun/\eta_s,      \label{eq:calpha} \\  
\mathrm{C}_{BL}&=&S_{BL_0}R_\sun/\eta_s, \label{eq:cbl} \\
\mbox{the ratio }&&\eta_r/\eta_s, \label{eq:dp} \\
\mbox{and the ratio }&&\eta_{cz}/\eta_s. \label{eq:dt}
\end{eqnarray}
The magnetic Reynolds number $\mathrm{Rm}$ is associated with the amplitude of the large-scale
meridional flow, $u_0$. The three following coefficients $\mathrm{C}_{\Omega}$, $\mathrm{C}_{\alpha}$, $\mathrm{C}_{BL}$
respectively express the ratio of the equatorial rotation, turbulent and Babcock-Leighton time 
scales to the diffusive time scale. In these expressions,  $\Omega_{eq}$ is the equatorial rotation rate, and $\alpha_0$ and $S_{BL_0}$ are the amplitudes of the $\alpha$ and 
BL terms seen above. 
The remaining two terms $\eta_r/\eta_s$ and $\eta_{cz}/\eta_s$ are magnetic diffusivity ratios entering the nondimensional 
forms of equations (\ref{diffP}) and (\ref{diffT}). The $\sim$ in Eqs. (\ref{induction_pol}) and (\ref{induction_tor}) denotes normalization with respect to those quantities. Note 
 that a suitable rescaling of $A_{\varphi}$ can decrease 
the number of control parameters by one, as it can scale either $\mathrm{C}_{\alpha}$ or $\mathrm{C}_{BL}$ out 
of the problem (it is the ratio of these two that would remain). 
Albeit more elegant, we did not consider this possibility. We shall
therefore analyse the $\alpha$ and BL effects independently in the remainder of this study.

\par
Finally, our formulation has to be complemented with boundary and initial conditions. The inner boundary condition is that of  a
perfect conductor. An approximation of this condition is that 
\begin{eqnarray}
A_{\varphi}=B_{\varphi}=0 \mbox{ at the inner radius}\,\, r=0.6R_\sun  \mbox{ \citep{chatterjee2004full}}.
\end{eqnarray}
The outer boundary condition corresponds to the interface with an insulating medium, and requires matching of the internal solar field
 with a potential field \citep{dikpati1999babcock}. 
 
 As
an initial condition, we choose a dipolar field confined inside the convection zone. 
In this case, 
\begin{eqnarray}
A_{\varphi}(\mathbf{r},t=0)&=&   \sin\theta/r^2 \mbox{ for }  0.7 R_{\sun} \leq r \leq  R_{\sun},\\ 
A_{\varphi}(\mathbf{r},t=0)&=& 0 \mbox{ elsewhere, } \\
B_{\varphi}(\mathbf{r},t=0)&=& 0  \mbox{ everywhere}.
\end{eqnarray}


The numerical approximation of the problem at hand is based on the 
Parody code, which was originally designed for three-dimensional 
geodynamo simulations  \citep{dormy1998mhd,aubert2008magnetic}, and successfully 
passed the dynamo benchmark of \citet{christensen2001numerical}. The magnetic field 
is expanded according to the three-dimensional poloidal-toroidal 
decomposition
\begin{eqnarray} \label{poltor_B} 
\mathbf{B}=\boldsymbol{\nabla}\times\boldsymbol{\nabla}\times(
\mathcal{P}\mathbf{r})
          +\boldsymbol{\nabla}\times(\mathcal{T}\mathbf{r}),
\end{eqnarray}
where the poloidal and toroidal scalar potentials 
$\mathcal{P}$ and $\mathcal{T}$ are further expanded 
upon an axisymmetric spherical harmonic basis $Y_{n}^0 (\theta)$,  
 according to 
\begin{eqnarray} \label{B_sh}
({\mathcal P},{\mathcal T})(r,\theta,t)
= \sum_{n = 1}^{N} ({\mathcal P}_n, {\mathcal T}_n)(r,t) \;  Y_{n}^0 (\theta), 
\end{eqnarray}
and truncated at spherical harmonic degree $N$. 
The discretization is completed by applying a second-order finite
differencing in radius and second order time integration, 
 comprising a Crank-Nicolson scheme for the diffusive terms and
 a second order Adams-Bashforth scheme for the nonlinear terms. 
The resulting code was then successfully tested against the reference
    solutions of \citet{jouve2008solar}. Details of this
 benchmark are provided in  Appendix \ref{appendix_1}.
  The results presented in what follows were obtained
   using $N=65$, and $N_r=65$ uniform radial levels in 
   $[0.6R_\sun,R_\sun]$, and a constant
    non-dimensional time step size $\Delta t=5\ 10^{-6}$.
 A typical run comprised $10^7$ time steps, which corresponds
  roughly to $25,000$~years. 

\section{Forward modelling: model properties and variability} \label{section_sensitivity}
With our operational purpose in mind, a representative solution of the model should match some of the basic solar cycle features 
\citep{charbonneau2010dynamo}: cyclic polarity reversals with approximately 11~years periodicity; strong toroidal fields at the base of the convection zone migrating from mid-latitudes
towards the equator; poleward migration of a weaker high-latitude magnetic field; phase lag of 
$\pi/2$ between the toroidal field at mid-latitudes and polar field at the poles; antisymmetry 
of the magnetic field between the North and South hemispheres; and long-term variability of the
solar cycle.

\par
In the following, we will impose the fixity of some of those non-dimensional numbers appearing in Eqs.~(\ref{eq:Rm})-(\ref{eq:dt}). As helioseismological data give $\Omega_{eq}\sim2\pi\times460.7$~nHz, we set accordingly $\mathrm{C}_{\Omega}=4.7~10^4$. 
 In addition, the turbulent diffusivity in the solar interior is not well constrained \citep{ossendrijver2003solar}, and
  we consequently hold for simplicity the ratios $\eta_r/\eta_s$ and $\eta_{cz}/\eta_s$ fixed to values (see Table~\ref{table_symb}) 
   previously shown to yield a satisfactory degree of solar semblance \citep[e.g.][]{dikpati1999babcock}. Variations of the remaining
free parameters $\mathrm{C}_{\alpha}$, $\mathrm{C}_{BL}$ and $\mathrm{Rm}$  allow for a broad range of solutions. $\mathrm{Rm}$  
represents the strength of the meridional circulation 
 and controls the periodicity of the
 solar cycle, a well-known characteristic of flux transport
dynamos \citep{dikpati1999babcock}. Consequently, and because 
of the strong observational constraint to obtain a period close
 to $11$~years, our family of models works in  relatively 
narrow range of $\mathrm{Rm}$. As the meridional flow measured at the solar surface at mid-latitudes has
an average magnitude $u_0$ of $15$~m/s \citep{hathaway1996doppler}, 
 we vary $\mathrm{Rm}$ between 308 and 378 ($u_0\approx13$ and $17$~m/s, respectively). 
Within this range, getting a self-sustained reversing dynamo
requires $\mathrm{C}_{\alpha}\gtrsim 2$ and $\mathrm{C}_{BL}\gtrsim 0.5$.

\par
\begin{figure}[h]
\epsscale{0.85}
\plotone{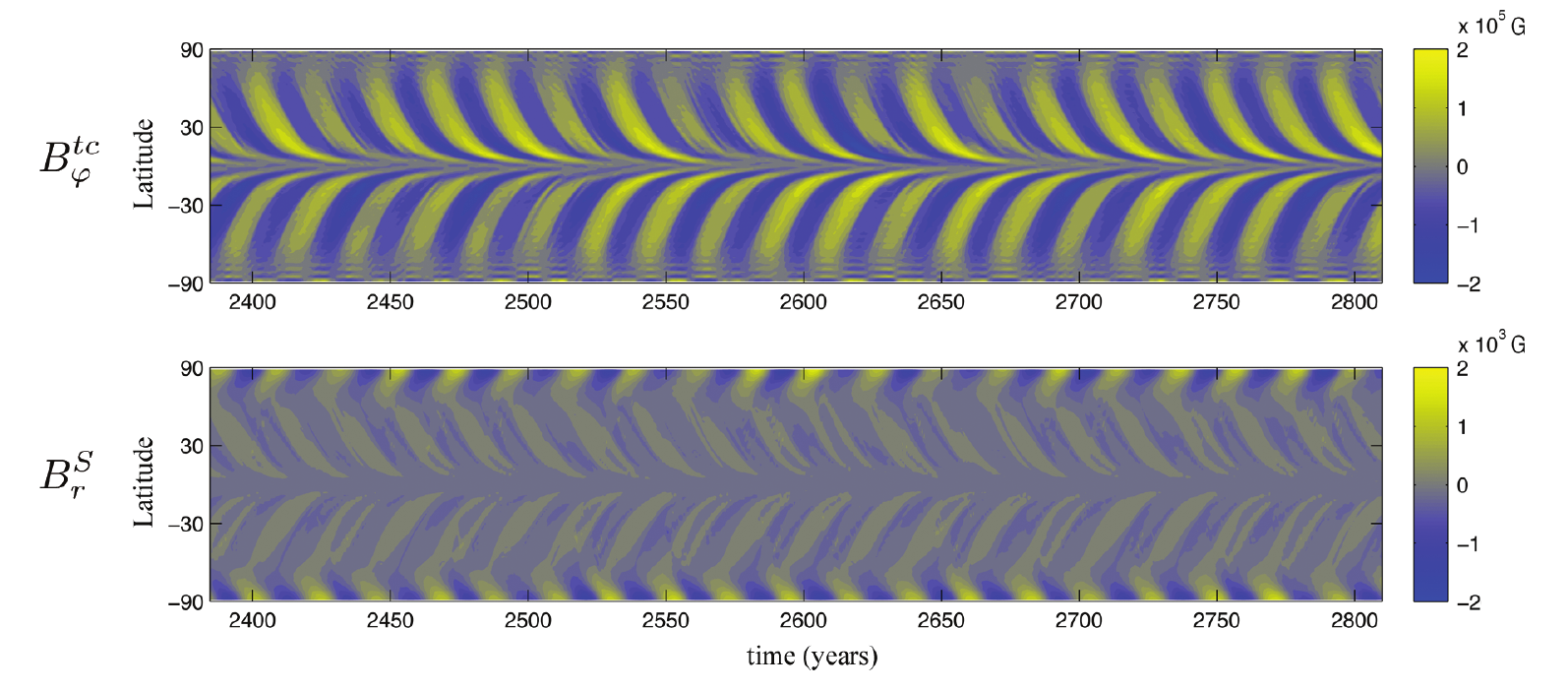}
\caption{Time-latitude (butterfly) diagrams of the reference solution $ S$, corresponding to $\mathrm{Rm}=318$, $\mathrm{C}_{\alpha}=8$ and $\mathrm{C}_{BL}=1$. Top: toroidal magnetic field at the tachocline; bottom: 
radial magnetic field at the solar surface.}
\label{fig_btfly}
\end{figure}

  
We pick a reference (standard) solution (labelled S in the following) which has $\mathrm{Rm}~=~318$, 
$\mathrm{C}_{\alpha}=8$ and $\mathrm{C}_{BL}=1$; it generates quasi-periodic reversals, separated by  approximately $10.95$~years. Figure~\ref{fig_btfly} represents 
the simulated evolutions of the toroidal field at the tachocline, $B_{\phi}^{tc}$, 
and of the radial field at the surface, $B_r^S$. It illustrates that
 the criteria for solar semblance which we listed are essentially met.
 This does not include the equatorial antisymmetric field configuration, a known 
  recurring issue with Babcock-Leighton models \citep{chatterjee2004full,charbonneau2010dynamo}. In this respect, 
  \cite{dikpati2001flux} and \cite{bonanno2002parity}
 previously showed that 
 the addition of an $\alpha$-effect in a thin layer above the tachocline
 (as done here, recall Fig.~\ref{fig_flow}(\textit{c}))
   helps in obtaining antisymmetric solutions. However, and even if the portion of the dynamical 
   trajectory represented in Fig.~\ref{fig_btfly} does display
    an antisymmetric magnetic field configuration, let us stress
     that there does not seem to exist a clear preferred mode of operation
   for the magnetic field: periods of symmetric, antisymmetric
    and out-of-phase modes alternate over the dynamical trajectory
    followed by the standard model. 
    
    Long-term variability of the solar cycle is also present in this reference solution.  \citet{charbonneau2005fluctuations} point out that chaotic modulation is a characteristic of 
Babcock-Leighton models in which the BL term includes a lower operational threshold, 
as is the case in our model. Short periods of weaker than average activity level, lasting for approximately 3~cycles, are frequently found in our simulations, 
 over a vast range of input parameters. In addition to this short-term variability, 
some of the solutions we obtain (including the reference solution S) display 
as well long periods of grand minima, lasting for several centuries, 
during which the cycle is not fully developed, 
but persists with a residual activity 
(see~\citeauthor{sanchezrevision}, in press, for more details). 
The occurrence of long periods of minimum activity is rare 
in our simulations; we chose accordingly to focus on their regular, quasi-cyclic behaviour to carry out the predictability analysis exposed below.

The quantities $\mathrm{C}_{\alpha}$ and
$\mathrm{C}_{BL}$ (recall their definition
 in Eqs.~(\ref{eq:calpha}) and
(\ref{eq:cbl})) are less tightly 
constrained by observations than $\mathrm{Rm}$, and they will  
 constitute the effective degrees of freedom of our class of models 
  when we investigate its horizon of predictability in the following section.
Variations in $\mathrm{C}_{\alpha}$ and $\mathrm{C}_{BL}$ affect the overall morphology of the solar cycle in 
different ways. 
While an increase in $\mathrm{C}_{\alpha}$ tends to excite higher frequencies during the solar cycle, it does not result in strong alterations of the magnetic field strength and cycle
periodicity. On the other hand, the intensity of the magnetic field is strongly and irregularly sensitive to 
variations of $\mathrm{C}_{BL}$ -- the overall trend is that it grows with 
$\mathrm{C}_{BL}$. Increasing $\mathrm{C}_{BL}$ also usually results in the appearance of a
feature respecting the Gnevyshev-Ohl rule, that is the persistent pattern of alternating high 
and low amplitudes of the solar cycles \citep{hathaway2010solar}. A too large an increase, 
 though, gives rise to intermittent, non solar-like, solutions. This forces us to define an 
 upper bound of $2$ for any admissible $\mathrm{C}_{BL}$. 
 On the other hand, as the main role 
 of $\mathrm{Rm}$ is that of setting the pace of 
 the solar cycle, increasing its value 
 leads to a shortening of the simulated periodicity (note that the first
 columns of Table~\ref{table_ts_all}, which we will discuss further below, 
  document in detail this variability).
 


\section{Predictability analysis} \label{section_eg}

\subsection{Methodology}

Our mean-field solar dynamo model is a dynamical system, characterized by a limited 
range of predictability, owing {\bf to} its chaotic nature \citep{lorenz1963deterministic}.
As stated in the introduction, two initially very close, {\sl twin}, dynamical trajectories
are bound to diverge in a finite time~$\tau$. The analysis of the divergence
between these twin trajectories forms the backbone of our methodology; it
is based on 
the work carried out by \cite{hulot2010earth} and \cite{lhuillier2011earth} to study the
 limit of predictability of the geodynamo.

We create a twin from a reference trajectory by perturbing 
a field variable (or control parameter) $\xi$ at a
given instant $t_p$ in the following way 
\begin{eqnarray} \label{perturb}
\xi (t_p) \longmapsto
\widetilde{\xi}(t_p)=\xi(t_p)(1+\varepsilon),
\end{eqnarray}
where $\widetilde{\xi}$ and 
$\varepsilon$
are the perturbed quantity and 
the amplitude 
of the perturbation, respectively. 

Of importance for the assessment of the predictability is the evolution of the
distance between the two trajectories over time. In order to monitor
this distance, we resort to two pointwise measures, which are 
related
 to the toroidal field $B_\varphi$ at a 
 point $\mathbf{r}_{tc}\equiv(r=r_{tc},\theta=70^{\mbox{\scriptsize o}})$ 
 on the tachocline, and to the 
 radial field $B_r$ at 
 a point $\mathbf{r}_{S}\equiv(r=R_{\sun},\theta\sim2^{\mbox{\scriptsize o}})$ at 
 the solar surface. 
 These measures write
\begin{eqnarray} \label{error_growtht}
\Delta B_\varphi(\mathbf{r}_{tc},t)\equiv 
\frac{|   B_\varphi(\mathbf{r}_{tc},t)
- \widetilde{B}_\varphi(\mathbf{r}_{tc},t)
|}%
{\sqrt{\langle B_\varphi^2(\mathbf{r}_{tc})\rangle}}
\end{eqnarray}
and
\begin{eqnarray} \label{error_growthr}
\Delta B_r(\mathbf{r}_{S},t)\equiv 
\frac{|   B_r(\mathbf{r}_{S},t)
- \widetilde{B}_r(\mathbf{r}_{S},t)
|}%
{\sqrt{\langle B_r^2(\mathbf{r}_{S})\rangle}}, 
\end{eqnarray}
respectively. In these two definitions, notice that the distance is normalized since 
the brackets $\langle \cdot \rangle$ represent time averaging (which we 
perform over a period of about $1,000$ years after $t=t_p$). In the following,
we will use $\Delta$ as a shorthand for
 $\Delta B_\varphi$ or $\Delta B_r$, when the distinction need not be made, 
 and we will refer to the evolution of $\Delta$
as the \textit{error growth}: in a forecasting perspective,
the perturbation which we insert can indeed be interpreted as the uncertainty
 affecting the initial condition (or the control parameters) of the model. In this
 sense the distance we measure is analogous to the growth of the 
 forecast error of interest for the data assimilation practitioner. 

\begin{figure}[h]
\epsscale{0.6}
\plotone{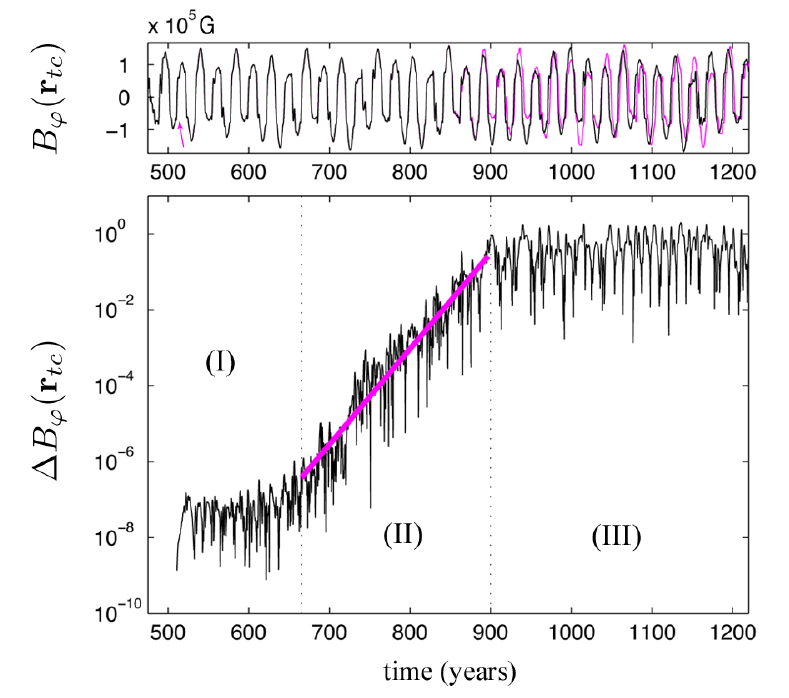}
\caption{Error growth behavior for the reference model S (see Table~\ref{table_ts_all}). Top: 
toroidal field at $20^0$ latitude on the tachocline. The reference solution (shown in black) is 
perturbed at a given instant, shown by the pink arrow,  by a relative amount $\varepsilon=10^{-6}$. 
This generates the perturbed solution (purple) which progressively diverges 
from the reference one.  
Bottom: the difference $\Delta B_\varphi$ between both solutions 
on a logarithmic scale. The error growth can be separated in 
three well-defined stages: (I) a mobilization phase, (II) an exponential growth 
phase and
(III) a saturated phase. In phase (II), we perform a least-squares 
regression (purple line) in order to estimate the error growth rate $\lambda$. See text for details.
}
\label{fig_egr}
\end{figure}

 Figure \ref{fig_egr} shows the typical evolution of the error growth (measured here 
 in terms of $\Delta B_\varphi$) in our numerical
  experiments. It corresponds to a $\varepsilon=10^{-6}$ perturbation
  applied to the spectral poloidal coefficient ${\mathcal P}_1$.  
  The evolution of $\Delta$ comprises three distinct phases. First, both trajectories remain
   fraternal, as their distance remain similar to $\varepsilon$ (phase I in  Fig.~\ref{fig_egr}). This is called
    the mobilization phase by \cite{lhuillier2011earth}. Next, 
the error enters a phase of exponential growth (phase II), until it reaches saturation (phase III). 
 From then on, the reference and perturbed solutions evolve in an uncorrelated way.

Among these three phases, the phase of exponential growth is the most meaningful to
constrain the limit of predictability. Considering that this phase starts at $t=t_\varepsilon$ with
 an initial value $\varepsilon$, the distance evolves according to 
\begin{eqnarray} 
\Delta(t)=\varepsilon\ e^{\lambda (t-t_\varepsilon)},
\label{eq:eg}
\end{eqnarray}
where $\lambda$ denotes the exponential growth rate. Its inverse 
$\lambda^{-1}$
is the so-called $e$-folding time $\tau_e$, namely the divergence time $\tau$ we discussed above.  
We set out to estimate $\lambda$ (or $\tau_e$) as accurately as possible for the 
class of mean-field models considered in this work. 
Visual inspection
of the time series of $\Delta$ allows us to pick the phase
of exponential growth; we next perform a least-squares analysis to estimate 
$\lambda$ (this procedure yields the purple line in 
 Fig.~\ref{fig_egr}). 
 
 That estimate may depend on the type and amplitude of the perturbation, though, which calls for
a systematic approach to evaluating $\lambda$. In the next
subsection, we use the standard model S presented in section~\ref{section_sensitivity}
to vary extensively the type and amplitudes of perturbations. 
Within this single-model context, we find that
 the characteristics of the error growth are robust. Therefore, in order
  to push the analysis further, we shall consider in section~\ref{sec:egvar} how   
 $\lambda$ may be influenced by the values of 
 the triplet~$(\mathrm{Rm},\mathrm{C}_{BL},\mathrm{C}_\alpha)$. 
\subsection{Error growth in the standard model}
\label{sec:egstd}

\subsubsection{Magnetic perturbations}

As explained above, we focus here on the standard model S and begin by examining its response
to perturbations of the magnetic field. We study different scenarios. The perturbation 
can affect either the poloidal scalar ${\mathcal P}$ or the toroidal scalar ${\mathcal T}$. It can be either large-scale (restricted to the $n=1$ harmonic degree), in which case it writes
\begin{eqnarray} \label{pert_dip}
{\mathcal P}_1(r,t_p) 
\longmapsto
\widetilde{{\mathcal P}_1}(r,t_p) ={\mathcal P}_1(r,t_p)(1+\varepsilon),
\end{eqnarray}
(and the same for $\mathcal{T}_1$), 
or distributed randomly over the
entire spectrum, according to
\begin{eqnarray} \label{pert_rand}
{\mathcal P}_n \longmapsto 
\widetilde{{\mathcal P}_n}(r,t_p)={\mathcal P}_n(r,t_p)(1+\gamma_n\varepsilon), 
\ 
1 \leq  n \leq N
,
\end{eqnarray}
(and the same for ${\mathcal T}_n$), 
in which the $\gamma_n$ are random numbers from $0$ to $1$ distributed over 
all the harmonic degrees. In the remainder of this subsection, the amplitude of the perturbation $\varepsilon$ is set to $10^{-6}$. 

\begin{figure}[h]
\epsscale{0.6}
\plotone{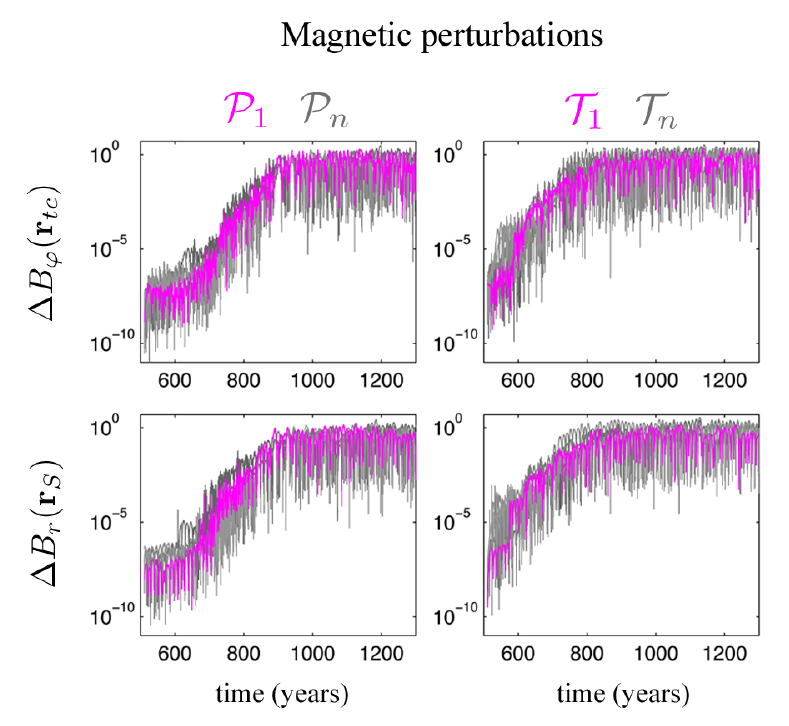}
\caption{Time series of $\Delta B_\varphi$ and $\Delta B_r$, following
the application of magnetic perturbations of relative amplitude $\varepsilon=10^{-6}$
on standard model S.
The perturbations are inserted either on the poloidal (left column) or 
toroidal (right column) component of the magnetic field, and they affect either 
the first harmonic degree 
($\mathcal{P}_1$/$\mathcal{T}_1$ pink curves) or  all
the harmonic degrees randomly ($\mathcal{P}_n$/$\mathcal{T}_n$), in which case
$5$ curves with different shades of grey are shown in each graph.}
\label{fig_case1}
\end{figure}

The pink curves in Figure~\ref{fig_case1} show that large scale perturbations of the poloidal
or toroidal scalars defined by Eq.~(\ref{pert_dip}) yield the same well-defined three phases for the evolution 
of $\Delta B_r$ and $\Delta B_\varphi$.  In addition, each panel of
 Fig.~\ref{fig_case1} comprises 5 grey curves obtained from
  5 random realisations of the small scale perturbations defined by Eq.~(\ref{pert_rand}). Despite some scatter, visual inspection indicates a common
error growth behaviour. In particular, if we were to estimate $\lambda$ from this catalog
of curves, we would probably get a robust value. This is rather encouraging, but 
before proceeding with the actual calculation of $\lambda$,  
 let us now inspect in more detail its sensitivity to a broader range of perturbations.
 
\subsubsection{Systematic perturbations}

\label{sec:syst_pert}

We thus investigate now the error growth 
induced by perturbations of different origins, varying amplitudes $\varepsilon$ 
and different times of insertion $t_p$ on the standard model S. The origin of the perturbation $\xi$ in Eq.~(\ref{perturb}) 
 can be one of the following: ${\mathcal P}_1$,
 $\mathcal{T}_1$, $\mathcal{P}_n$, $\mathcal{T}_n$ (as in the previous section), $\Omega$ or $u_p$ (the flow), $\alpha$ or $S_{BL}$ (the poloidal source terms), or $\eta_p$ (the  poloidal diffusivity profile).  
For each of these $9$ possibilities, we consider perturbations of amplitudes $10^{-2}$, $10^{-4}$, $10^{-6}$, $10^{-8}$ or 
$10^{-10}$. Finally, we perturb the reference dynamical trajectory at three different times, $t_p = t_1, 
t_2 $ or $t_3$. We therefore consider 
9~(origins)~$\times$~5~(amplitudes)~$\times$~3~(instants)$~=~135$~ways of
perturbing the standard trajectory. 
 Since both $\Delta B_\varphi$ and $\Delta B_r$ are used to monitor the error growth, this allows us to construct a database of $270$~estimates of $\lambda$. The database is completely described in Table~\ref{table_lambda_ref}. 
 
\begin{landscape}
\begin{table}
\begin{center}
\scriptsize
\rotate
\caption{Mean values of $\lambda$ and its uncertainties $\sigma$ (in units of $10^{-2}$ years$^{-1}$) from the systematic perturbation analysis of the standard model S.}
\begin{tabular}{cr cccccccccc|cccccccccc}
\tableline \tableline
& & \multicolumn{10}{c}{$\Delta B_{\varphi}$} & \multicolumn{10}{c}{$\Delta B_{r}$} \\
&
$\varepsilon$  &
\multicolumn{2}{c}{$10^{-10}$} & \multicolumn{2}{c}{$10^{-8}$} &
\multicolumn{2}{c}{$10^{-6}$} & \multicolumn{2}{c}{$10^{-4}$} & \multicolumn{2}{c}{$10^{-2}$} &
\multicolumn{2}{c}{$10^{-10}$} & \multicolumn{2}{c}{$10^{-8}$} &
\multicolumn{2}{c}{$10^{-6}$} & \multicolumn{2}{c}{$10^{-4}$} & \multicolumn{2}{c}{$10^{-2}$} \\

 $t_p$ &
$\xi$ &
$\lambda$ & $\sigma$ & $\lambda$ & $\sigma$ & $\lambda$ & $\sigma$ & $\lambda$ & $\sigma$
& $\lambda$ & $\sigma$ &
$\lambda$ & $\sigma$ & $\lambda$ & $\sigma$ & $\lambda$ & $\sigma$ & $\lambda$ & $\sigma$
& $\lambda$ & $\sigma$ \\

\tableline

$t_1$ & $\mathcal{P}_1$ & $ 3.66 $ & $ 0.07 $ & $ 2.49 $ & $ 0.05 $ & $ 2.52 $ & $ 0.06 $ & $ 1.68 $ & $ 0.09 $ & $ 3.57 $ & $ 0.21 $ & $ 3.69 $ & $ 0.09 $ & $ 2.33 $ & $ 0.06 $ & $ 2.58 $ & $ 0.06 $ & $ 1.44 $ & $ 0.09 $ & $ 4.23 $ & $ 0.35 $  \\
& $\mathcal{T}_1$ & $ 3.62 $ & $ 0.07 $ & $ 3.20 $ & $ 0.13 $ & $ 1.52 $ & $ 0.05 $ & $ 1.33 $ & $ 0.07 $ & $ 3.82 $ & $ 0.35 $ & $ 3.70 $ & $ 0.09 $ & $ 3.51 $ & $ 0.11 $ & $ 1.51 $ & $ 0.05 $ & $ 1.60 $ & $ 0.09 $ & $ 1.19 $ & $ 0.22 $  \\
& $\mathcal{P}_n$ & $ 3.48 $ & $ 0.06 $ & $ 1.49 $ & $ 0.06 $ & $ 1.87 $ & $ 0.08 $ & $ 1.21 $ & $ 0.08 $ & $ 2.76 $ & $ 0.27 $ & $ 3.69 $ & $ 0.08 $ & $ 1.51 $ & $ 0.05 $ & $ 2.05 $ & $ 0.06 $ & $ 1.33 $ & $ 0.07 $ & $ 1.21 $ & $ 0.21 $  \\
& $\mathcal{T}_n$ & $ 2.89 $ & $ 0.06 $ & $ 2.12 $ & $ 0.08 $ & $ 1.86 $ & $ 0.11 $ & $ 2.43 $ & $ 0.21 $ & $ 1.61 $ & $ 0.15 $ & $ 3.04 $ & $ 0.06 $ & $ 2.11 $ & $ 0.07 $ & $ 1.89 $ & $ 0.10 $ & $ 2.29 $ & $ 0.14 $ & $ 1.52 $ & $ 0.13 $  \\
& $\Omega$ & $ 3.16 $ & $ 0.05 $ & $ 3.82 $ & $ 0.09 $ & $ 2.30 $ & $ 0.06 $ & $ 1.40 $ & $ 0.05 $ & $ 2.11 $ & $ 0.09 $ & $ 3.98 $ & $ 0.10 $ & $ 3.52 $ & $ 0.11 $ & $ 2.90 $ & $ 0.08 $ & $ 2.22 $ & $ 0.06 $ & $ 2.20 $ & $ 0.14 $  \\
& $u_p$ & $ 1.74 $ & $ 0.03 $ & $ 3.52 $ & $ 0.04 $ & $ 2.90 $ & $ 0.08 $ & $ 3.98 $ & $ 0.06 $ & $ 2.20 $ & $ 0.14 $ & $ 2.57 $ & $ 0.05 $ & $ 3.73 $ & $ 0.07 $ & $ 2.77 $ & $ 0.07 $ & $ 2.43 $ & $ 0.06 $ & $ 2.15 $ & $ 0.13 $  \\
& $\alpha$ & $ 3.12 $ & $ 0.05 $ & $ 2.32 $ & $ 0.04 $ & $ 3.03 $ & $ 0.05 $ & $ 3.98 $ & $ 0.12 $ & $ 2.01 $ & $ 0.05 $ & $ 3.18 $ & $ 0.05 $ & $ 2.28 $ & $ 0.03 $ & $ 3.39 $ & $ 0.05 $ & $ 4.19 $ & $ 0.15 $ & $ 1.62 $ & $ 0.06 $  \\
& $S_{BL}$ & $ 2.73 $ & $ 0.05 $ & $ 3.39 $ & $ 0.08 $ & $ 1.92 $ & $ 0.04 $ & $ 1.91 $ & $ 0.13 $ & $ 4.83 $ & $ 0.25 $ & $ 2.91 $ & $ 0.05 $ & $ 3.58 $ & $ 0.09 $ & $ 1.84 $ & $ 0.05 $ & $ 1.67 $ & $ 0.09 $ & $ 2.18 $ & $ 0.14 $  \\
& $\eta_p$ & $ 2.66 $ & $ 0.05 $ & $ 2.53 $ & $ 0.05 $ & $ 1.83 $ & $ 0.04 $ & $ 2.22 $ & $ 0.11 $ & $ 1.56 $ & $ 0.28 $ & $ 2.90 $ & $ 0.04 $ & $ 2.31 $ & $ 0.06 $ & $ 1.85 $ & $ 0.05 $ & $ 1.12 $ & $ 0.08 $ & $ 1.12 $ & $ 0.12 $  \\
\tableline
$t_2$ & $\mathcal{P}_1$ & $ 2.48 $ & $ 0.04 $ & $ 2.57 $ & $ 0.04 $ & $ 1.37 $ & $ 0.06 $ & $ 1.46 $ & $ 0.04 $ & $ 1.37 $ & $ 0.07 $ & $ 2.50 $ & $ 0.04 $ & $ 2.77 $ & $ 0.05 $ & $ 1.26 $ & $ 0.05 $ & $ 1.15 $ & $ 0.05 $ & $ 1.61 $ & $ 0.14 $  \\
& $\mathcal{T}_1$ & $ 2.62 $ & $ 0.04 $ & $ 3.69 $ & $ 0.06 $ & $ 3.15 $ & $ 0.10 $ & $ 1.36 $ & $ 0.04 $ & $ 1.39 $ & $ 0.09 $ & $ 2.76 $ & $ 0.04 $ & $ 3.97 $ & $ 0.07 $ & $ 3.91 $ & $ 0.12 $ & $ 1.28 $ & $ 0.07 $ & $ 0.87 $ & $ 0.16 $  \\
& $\mathcal{P}_n$ & $ 3.06 $ & $ 0.06 $ & $ 3.92 $ & $ 0.07 $ & $ 1.97 $ & $ 0.08 $ & $ 1.58 $ & $ 0.06 $ & $ 1.81 $ & $ 0.15 $ & $ 2.91 $ & $ 0.10 $ & $ 4.12 $ & $ 0.10 $ & $ 1.84 $ & $ 0.06 $ & $ 1.67 $ & $ 0.05 $ & $ 1.31 $ & $ 0.09 $  \\
& $\mathcal{T}_n$ & $ 3.44 $ & $ 0.06 $ & $ 2.87 $ & $ 0.08 $ & $ 2.07 $ & $ 0.08 $ & $ 1.53 $ & $ 0.08 $ & $ 2.10 $ & $ 0.23 $ & $ 3.61 $ & $ 0.08 $ & $ 2.76 $ & $ 0.08 $ & $ 2.48 $ & $ 0.11 $ & $ 1.37 $ & $ 0.06 $ & $ 1.79 $ & $ 0.22 $  \\
& $\Omega$ & $ 2.53 $ & $ 0.04 $ & $ 2.71 $ & $ 0.05 $ & $ 3.46 $ & $ 0.07 $ & $ 2.13 $ & $ 0.05 $ & $ 1.43 $ & $ 0.05 $ & $ 2.50 $ & $ 0.04 $ & $ 2.79 $ & $ 0.05 $ & $ 3.93 $ & $ 0.10 $ & $ 2.07 $ & $ 0.05 $ & $ 1.40 $ & $ 0.04 $  \\
& $u_p$ & $ 2.68 $ & $ 0.04 $ & $ 2.80 $ & $ 0.05 $ & $ 3.47 $ & $ 0.05 $ & $ 2.16 $ & $ 0.11 $ & $ 1.96 $ & $ 0.09 $ & $ 2.67 $ & $ 0.04 $ & $ 2.54 $ & $ 0.04 $ & $ 3.60 $ & $ 0.06 $ & $ 1.32 $ & $ 0.05 $ & $ 2.16 $ & $ 0.10 $  \\
& $\alpha$ & $ 3.11 $ & $ 0.05 $ & $ 2.81 $ & $ 0.05 $ & $ 4.03 $ & $ 0.04 $ & $ 2.65 $ & $ 0.05 $ & $ 3.35 $ & $ 0.09 $ & $ 3.19 $ & $ 0.05 $ & $ 2.83 $ & $ 0.06 $ & $ 2.31 $ & $ 0.04 $ & $ 3.07 $ & $ 0.05 $ & $ 2.57 $ & $ 0.11 $  \\
& $S_{BL}$ & $ 2.65 $ & $ 0.05 $ & $ 3.65 $ & $ 0.06 $ & $ 4.03 $ & $ 0.12 $ & $ 2.34 $ & $ 0.08 $ & $ 1.87 $ & $ 0.17 $ & $ 2.49 $ & $ 0.05 $ & $ 1.68 $ & $ 0.08 $ & $ 3.53 $ & $ 0.12 $ & $ 1.42 $ & $ 0.07 $ & $ 1.30 $ & $ 0.08 $  \\
& $\eta_p$ & $ 2.47 $ & $ 0.04 $ & $ 3.15 $ & $ 0.04 $ & $ 1.98 $ & $ 0.12 $ & $ 2.22 $ & $ 0.08 $ & $ 1.78 $ & $ 0.08 $ & $ 2.57 $ & $ 0.06 $ & $ 1.42 $ & $ 0.05 $ & $ 1.38 $ & $ 0.04 $ & $ 1.42 $ & $ 0.09 $ & $ 1.98 $ & $ 0.14 $  \\
\tableline
$t_3$ & $\mathcal{P}_1$ & $ 2.37 $ & $ 0.04 $ & $ 2.32 $ & $ 0.03 $ & $ 2.85 $ & $ 0.05 $ & $ 2.05 $ & $ 0.12 $ & $ 2.53 $ & $ 0.21 $ & $ 2.75 $ & $ 0.05 $ & $ 2.26 $ & $ 0.03 $ & $ 2.76 $ & $ 0.05 $ & $ 2.94 $ & $ 0.20 $ & $ 3.06 $ & $ 0.35 $  \\
& $\mathcal{T}_1$ & $ 2.59 $ & $ 0.05 $ & $ 2.57 $ & $ 0.07 $ & $ 3.19 $ & $ 0.07 $ & $ 2.45 $ & $ 0.16 $ & $ 1.37 $ & $ 0.32 $ & $ 2.69 $ & $ 0.06 $ & $ 2.07 $ & $ 0.06 $ & $ 2.61 $ & $ 0.08 $ & $ 3.65 $ & $ 0.25 $ & $ 1.84 $ & $ 0.22 $  \\
& $\mathcal{P}_n$ & $ 4.40 $ & $ 0.12 $ & $ 2.48 $ & $ 0.05 $ & $ 2.73 $ & $ 0.14 $ & $ 0.82 $ & $ 0.33 $ & $ 2.00 $ & $ 0.29 $ & $ 2.73 $ & $ 0.06 $ & $ 2.47 $ & $ 0.05 $ & $ 2.45 $ & $ 0.12 $ & $ 2.52 $ & $ 0.20 $ & $ 2.06 $ & $ 0.25 $  \\
& $\mathcal{T}_n$ & $ 1.64 $ & $ 0.02 $ & $ 2.81 $ & $ 0.06 $ & $ 2.81 $ & $ 0.10 $ & $ 3.07 $ & $ 0.19 $ & $ 2.72 $ & $ 0.32 $ & $ 1.96 $ & $ 0.04 $ & $ 2.59 $ & $ 0.05 $ & $ 2.48 $ & $ 0.10 $ & $ 3.16 $ & $ 0.24 $ & $ 3.65 $ & $ 0.34 $  \\
& $\Omega$ & $ 1.50 $ & $ 0.01 $ & $ 1.87 $ & $ 0.03 $ & $ 3.07 $ & $ 0.08 $ & $ 2.74 $ & $ 0.18 $ & $ 1.98 $ & $ 0.14 $ & $ 1.54 $ & $ 0.01 $ & $ 1.91 $ & $ 0.03 $ & $ 3.29 $ & $ 0.07 $ & $ 2.65 $ & $ 0.13 $ & $ 2.82 $ & $ 0.22 $  \\
& $u_p$ & $ 2.77 $ & $ 0.06 $ & $ 2.77 $ & $ 0.06 $ & $ 3.53 $ & $ 0.08 $ & $ 3.17 $ & $ 0.06 $ & $ 3.11 $ & $ 0.05 $ & $ 2.70 $ & $ 0.06 $ & $ 2.61 $ & $ 0.06 $ & $ 3.43 $ & $ 0.08 $ & $ 2.96 $ & $ 0.06 $ & $ 2.02 $ & $ 0.05 $  \\
& $\alpha$ & $ 1.76 $ & $ 0.03 $ & $ 3.56 $ & $ 0.09 $ & $ 1.33 $ & $ 0.04 $ & $ 2.10 $ & $ 0.04 $ & $ 2.95 $ & $ 0.05 $ & $ 1.67 $ & $ 0.03 $ & $ 2.66 $ & $ 0.06 $ & $ 1.60 $ & $ 0.02 $ & $ 2.15 $ & $ 0.04 $ & $ 2.57 $ & $ 0.06 $  \\
& $S_{BL}$ & $ 2.42 $ & $ 0.08 $ & $ 2.44 $ & $ 0.07 $ & $ 2.80 $ & $ 0.05 $ & $ 2.84 $ & $ 0.18 $ & $ 4.31 $ & $ 0.35 $ & $ 2.64 $ & $ 0.06 $ & $ 2.03 $ & $ 0.06 $ & $ 2.60 $ & $ 0.07 $ & $ 2.11 $ & $ 0.09 $ & $ 1.51 $ & $ 0.20 $  \\
& $\eta_p$ & $ 2.74 $ & $ 0.06 $ & $ 3.36 $ & $ 0.06 $ & $ 2.21 $ & $ 0.08 $ & $ 2.75 $ & $ 0.17 $ & $ 3.59 $ & $ 0.21 $ & $ 2.39 $ & $ 0.03 $ & $ 3.26 $ & $ 0.07 $ & $ 1.55 $ & $ 0.05 $ & $ 3.24 $ & $ 0.19 $ & $ 3.46 $ & $ 0.34 $  \\

\tableline
\end{tabular}
\label{table_lambda_ref}
\end{center}
\end{table}
\end{landscape}
 
\begin{figure}[ht!]
\epsscale{0.9}
\plotone{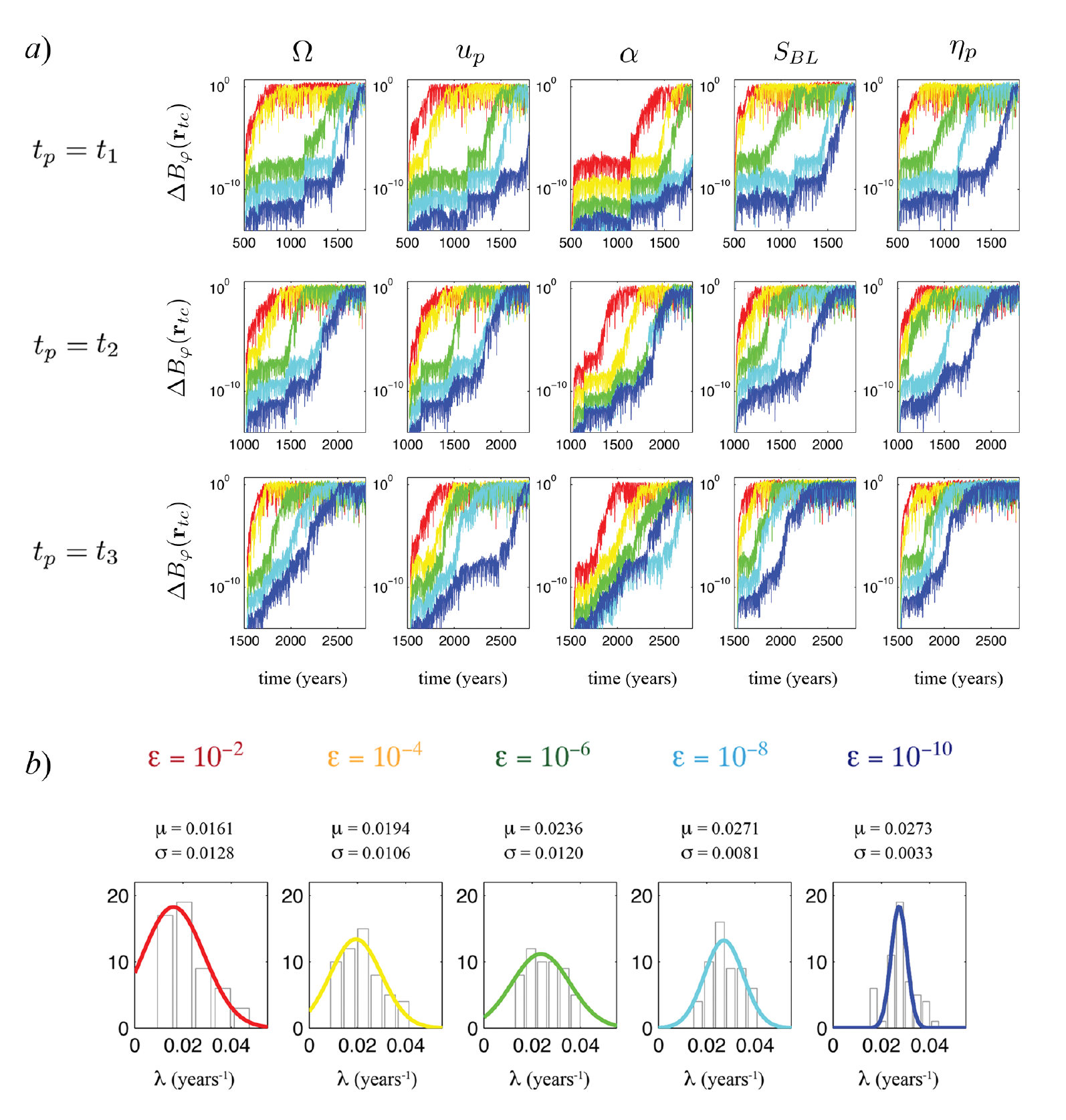}
\caption{Sensitivity of the error growth rate against perturbation types for model S: \textit{a}) Error growth considering different origins ($\xi~=~\Omega, u_p, \alpha, S_{BL}$ and $\eta_p$), times ($t_p~=~t_1,~t_2$ and $t_3$) and 
amplitudes ($\varepsilon=10^{-2},10^{-4},10^{-6},10^{-8}$ and $10^{-10}$); \textit{b}) 
histograms of the exponential growth rates $\lambda$ from the set of 
perturbations 
displayed in Table~\ref{table_lambda_ref} ordered by perturbation amplitude. The histograms 
are modeled by Gaussian curves with mean $\mu$ and standard deviation $\sigma$.}
\label{fig_t1rm2}
\end{figure}

Fig.~\ref{fig_t1rm2} means at illustrating the 
variability within the database of model S, and shows that 
regardless of this variability, the error growth displays a fair amount of
dynamical similarity in the $270$ scenarios we envisioned.
 Fig. \ref{fig_t1rm2}(\textit{a}) shows the evolution of
  $\Delta B_\varphi$, for different origins, times of perturbation insertion and different perturbation amplitudes. 
   We see that the error growth is weakly sensitive to 
   the origin of the perturbations. Still, the mobilization phase seems to vary depending on the way the perturbations were inserted. For perturbations corresponding to $\xi=\alpha$ or $\xi=u_p$, the mobilization phase lasts longer 
     (several centuries), and there is a mild dependency of the duration 
      of that phase on $t_p$.  The
   mobilization phase has a duration which decreases with 
    $\varepsilon$ as well. However, this variability on the mobilization phase does not strongly affect the estimate of $\lambda$.  On another note, it can also be seen that the error
growth due to smaller perturbations can experience 
secondary mobilization phases, and resume its exponential
growth after some time.

  Figure \ref{fig_t1rm2}(\textit{b}) presents the distribution of the 
  error growth rates (one histogram per value of $\varepsilon$, which 
  integrates all other dependencies) of model S. 
 The exponential growth is steeper for smaller
     levels of perturbations (noticing that $\varepsilon=10^{-8}$ and
      $\varepsilon=10^{-10}$ yield essentially the same behaviour, though), that is, large perturbations 
      lead to smaller values of~$\lambda$. There
       is also a general tendency for the growth to slacken as the error reaches macroscopic values.
  
   Still, focusing on small to extremely small 
   values of $\varepsilon$ ($10^{-6}$ and less), the estimated $\lambda$ does not vary by more than  $20$\%.  This robustness suggests the fact that $\lambda$ is an intrinsic property of our standard
    model S: regardless of the perturbation time and origin, and as long as it is small, 
     the exponential growth of the error is likely 
     to occur on a time scale $\tau$ of roughly $40$~years, that is over slightly more than three simulated
      cycles. 
      More precisely,  if 
      $T_c$ denotes the period of the simulated cycle
      and considering a least-squares analysis 
      of the $\varepsilon=10^{-10}$ histogram, we find that
       $\tau_e = \left(3.34 \pm 0.40\right)$~$T_c$. 

\subsection{Sensitivity of $\lambda$ to the control parameters}

\label{sec:egvar}

We now  explore the more general dependency of $\tau_e$
to the control parameters of our class of mean-field models. 
Since the simulated $T_c$ varies with these parameters as well, and since 
we wish to express $\tau_e$ in units of $T_c$, we investigate the joint
dependency of these two quantities on the triplet 
$(\mathrm{Rm},\mathrm{C}_\alpha,\mathrm{C}_{BL})$. 
\par
\begin{figure}[!hb]
\epsscale{0.9}
\plotone{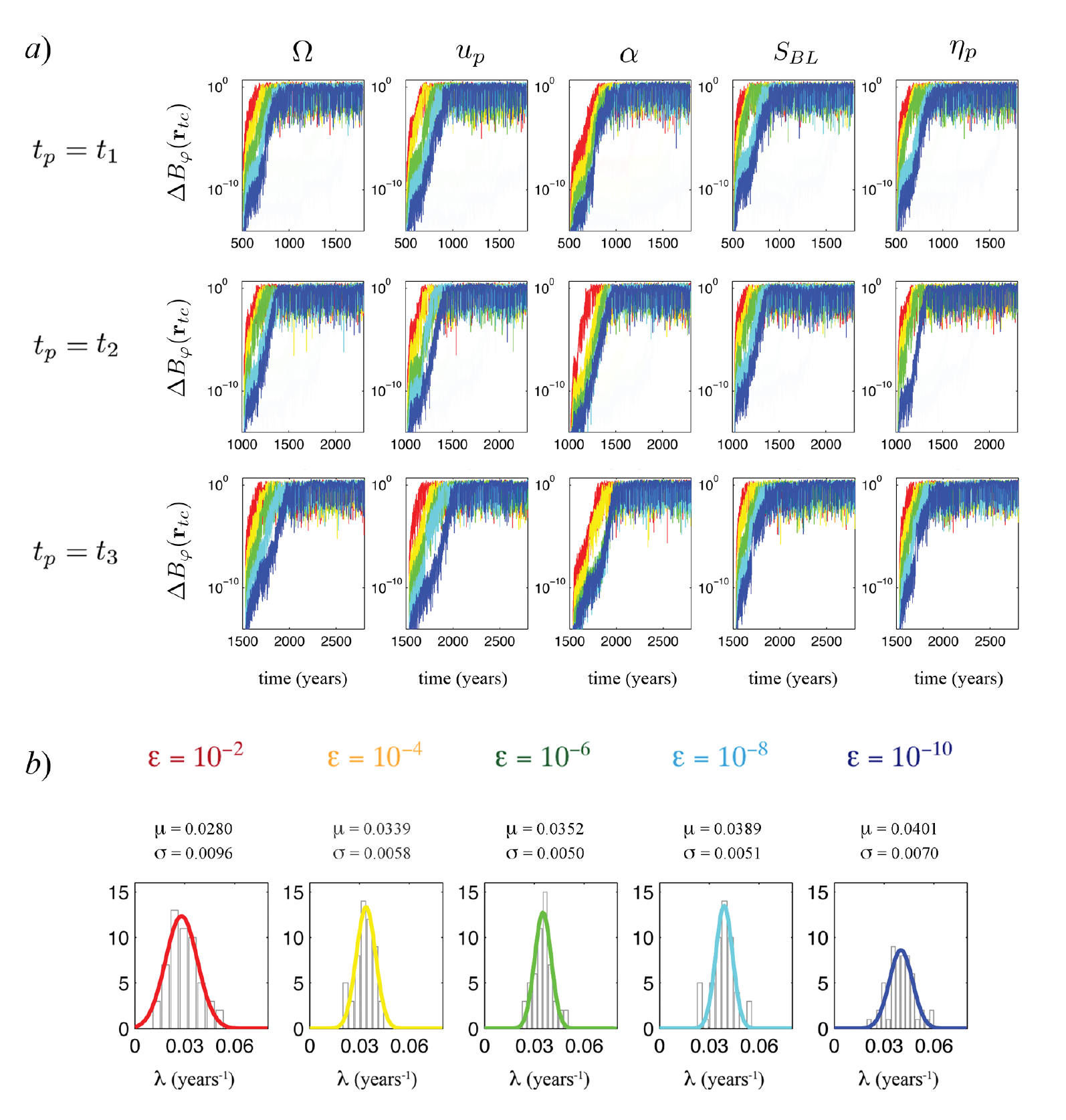}
\caption{Same as Figure \ref{fig_t1rm2}, for model T which has a 
stronger $\alpha$-effect than model S 
($\mathrm{C}_{\alpha}=16$ instead of $\mathrm{C}_{\alpha}=8$), all other control parameters
being the same.}
\label{fig_t2rm2}
\end{figure}

First, we increase the $\alpha$-effect coefficient from $\mathrm{C}_{\alpha}=8$ to 
$\mathrm{C}_{\alpha} = 16$ and consider the same $270$ possibilities as the ones
used for the standard model (this new model is labelled T in the following). Figure \ref{fig_t2rm2} illustrates
the corresponding database of model T, and highlights consistent differences when
compared with the standard case S shown in Fig.~\ref{fig_t1rm2}. 
Most notably, the mobilization phase is in every instance much shorter
 (not lasting more than a few decades), while the exponential growth phase
  is in all cases much steeper, two effects pointing towards an increased
   influence of turbulence as the value of $\mathrm{C}_{\alpha}$ increases, 
   leading to larger estimates for $\lambda$. We still
  retrieve the tendency for $\lambda$ to decrease with increasing $\varepsilon$, while
   its uncertainties decrease with $\varepsilon$. Accordingly, we find that 
    for $\varepsilon=10^{-10}$, $\tau_e = \left(2.45 \pm 0.42\right)$~$T_c$ (here, $T_c=10.15$~years).

\begin{figure}[h]
\epsscale{0.85}
\plotone{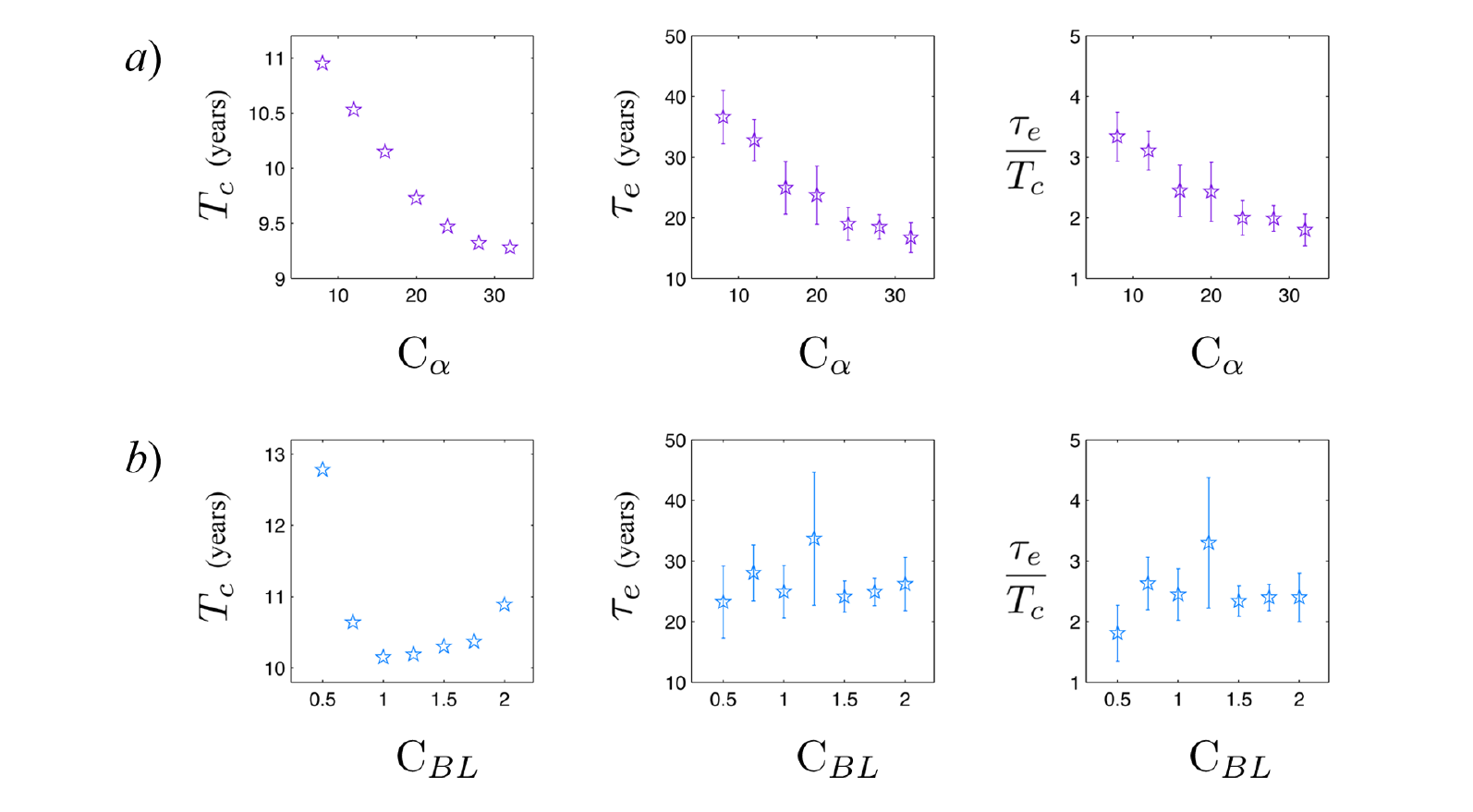}
\caption{
Solar cycle periodicity $T_c$, $e$-folding time $\tau_e$, and their ratio 
for a) (top row) $\mathrm{Rm}=318$, $\mathrm{C}_{BL}=1.0$ and a varying
   $\mathrm{C}_{\alpha}$ and b) (bottom row) $\mathrm{Rm}=318$, $\mathrm{C}_{\alpha}=16$, 
   and a varying $\mathrm{C }_{BL}$. 
}
\label{fig_time_scales_CaCs}
\end{figure}

\begin{figure}[h!]
\epsscale{0.85}
\plotone{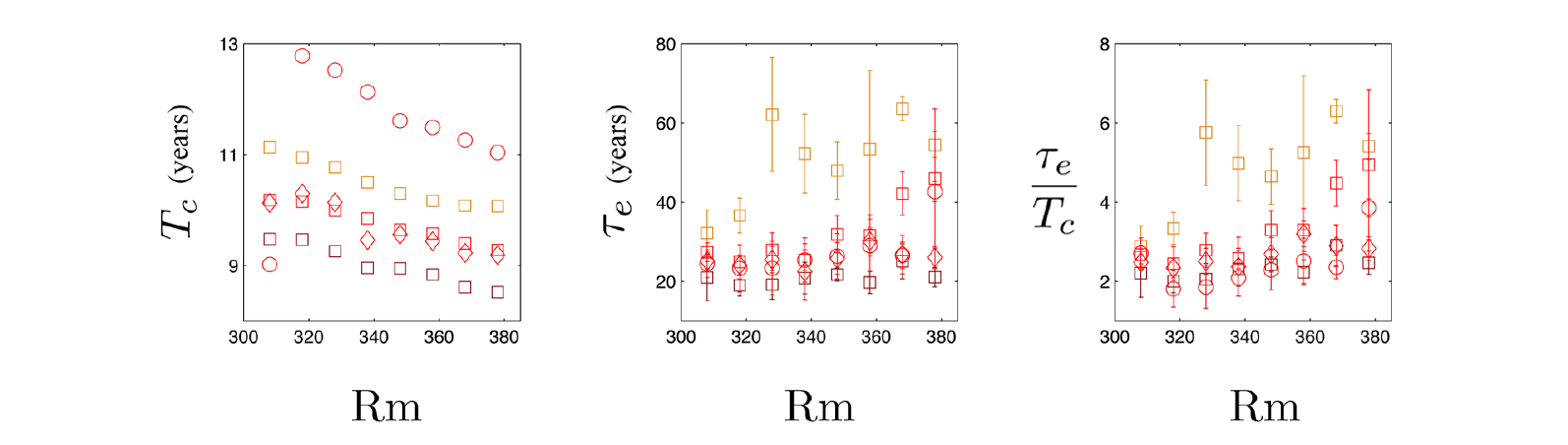}
\caption{Solar cycle periodicity $T_c$, $e$-folding time $\tau_e$ and their ratio 
for different values of $\mathrm{C}_{\alpha}$, $\mathrm{C}_{BL}$ and $\mathrm{Rm}$. 
The magnitude of $\mathrm{C}_{\alpha}$ is color-coded: $\mathrm{C}_{\alpha}=8$, orange; 
$\mathrm{C}_{\alpha}=16$, red; $\mathrm{C}_{\alpha}=24$, dark red. Symbols 
indicate different $\mathrm{C}_{BL}$: $\mathrm{C}_{BL}=0.5$, circles; 
$\mathrm{C}_{BL}=1.0$, squares; $\mathrm{C}_{BL}=1.5$, diamonds.}
\label{fig_time_scales_Rm}
\end{figure}

Next, we carry out a similar sensitivity
analysis with different triplets $(\mathrm{Rm},\mathrm{C}_\alpha,\mathrm{C}_{BL})$. More specifically, 
we consider the following possibilities 
\begin{itemize}
\item $308 \leq \mathrm{Rm} \leq 378 $,
\item  $0.5\leq \mathrm{C}_{BL} \leq 2$,
\item   $8 \leq \mathrm{C}_{\alpha}\leq 32$,
\end{itemize}
providing a total of $48$~different models (including models S and T). For each model we calculate $\lambda$
restraining the amplitude of the perturbation $\varepsilon$ to what we consider its most reliable level, 
namely $10^{-10}$. This survey is summarized in Table~\ref{table_ts_all}, and the
results (expressed in terms of the corresponding time scales $T_c$ and $\tau_e$) are shown in Figures~\ref{fig_time_scales_CaCs}-\ref{fig_time_scales_Rm}. 

\begin{table}[b!]
\begin{center}
\scriptsize
\caption{Summary of the values of solar cycle periodicity $T_c$, $e$-folding time $\tau_e$ and its uncertainty $\delta$, and the ratio $\tau_e/T_c$ for a large number of configurations of the triplet ($\mathrm{Rm}$,$\mathrm{C}_\alpha$,$\mathrm{C}_{BL}$). The letters S and T make reference to the main two models discussed in the bulk of the paper. All the time-scales are expressed in years.}
\begin{tabular} {rcrcrcrc|crcrcrc}
\tableline \tableline

& $ C_{BL} $ & $ C_{\alpha} $ & $ Rm $ & $ T_c $ & $ \tau_e $ & $ \delta$  & $ \tau_e/T_c $ &
$ C_{BL} $ & $ C_{\alpha} $ & $ Rm $ & $ T_c $ & $ \tau_e $ & $ \delta$ & $ \tau_e/T_c $\\

\tableline

& $ 0.50 $ & $ 16 $ & $ 308 $ & $ 9.02 $ & $ 24.48 $ & $ 3.50 $ & $ 2.71 $ & $ 1.00 $ & $ 16 $ & $ 368 $ & $ 9.40 $ & $ 42.15 $ & $ 5.50 $ & $ 4.48 $  \\
& $ 0.50 $ & $ 16 $ & $ 318 $ & $ 12.78 $ & $ 23.25 $ & $ 5.94 $ & $ 1.81 $ & $ 1.00 $ & $ 16 $ & $ 378 $ & $ 9.28 $ & $ 46.00 $ & $ 17.59 $ & $ 4.95 $  \\
& $ 0.50 $ & $ 16 $ & $ 328 $ & $ 12.52 $ & $ 23.33 $ & $ 6.79 $ & $ 1.86 $ & $ 1.00 $ & $ 20 $ & $ 318 $ & $ 9.73 $ & $ 23.74 $ & $ 4.77 $ & $ 2.43 $  \\
& $ 0.50 $ & $ 16 $ & $ 338 $ & $ 12.13 $ & $ 25.40 $ & $ 5.61 $ & $ 2.09 $ & $ 1.00 $ & $ 24 $ & $ 308 $ & $ 9.48 $ & $ 20.98 $ & $ 5.80 $ & $ 2.21 $  \\
& $ 0.50 $ & $ 16 $ & $ 348 $ & $ 11.61 $ & $ 26.37 $ & $ 5.74 $ & $ 2.29 $ & $ 1.00 $ & $ 24 $ & $ 318 $ & $ 9.47 $ & $ 19.00 $ & $ 2.72 $ & $ 2.00 $  \\
& $ 0.50 $ & $ 16 $ & $ 358 $ & $ 11.49 $ & $ 29.06 $ & $ 6.61 $ & $ 2.52 $ & $ 1.00 $ & $ 24 $ & $ 328 $ & $ 9.26 $ & $ 19.15 $ & $ 3.67 $ & $ 2.06 $  \\
& $ 0.50 $ & $ 16 $ & $ 368 $ & $ 11.26 $ & $ 26.60 $ & $ 3.29 $ & $ 2.36 $ & $ 1.00 $ & $ 24 $ & $ 338 $ & $ 9.94 $ & $ 20.85 $ & $ 4.02 $ & $ 2.09 $  \\
& $ 0.50 $ & $ 16 $ & $ 378 $ & $ 11.04 $ & $ 42.66 $ & $ 2.52 $ & $ 3.86 $ & $ 1.00 $ & $ 24 $ & $ 348 $ & $ 8.96 $ & $ 21.75 $ & $ 1.67 $ & $ 2.42 $  \\
& $ 0.75 $ & $ 16 $ & $ 318 $ & $ 10.64 $ & $ 28.05 $ & $ 4.62 $ & $ 2.63 $ & $ 1.00 $ & $ 24 $ & $ 358 $ & $ 8.84 $ & $ 19.74 $ & $ 2.81 $ & $ 2.23 $  \\
& $ 1.00 $ & $ 8 $ & $ 308 $ & $ 11.13 $ & $ 32.21 $ & $ 5.67 $ & $ 2.89 $ & $ 1.00 $ & $ 24 $ & $ 368 $ & $ 8.61 $ & $ 25.07 $ & $ 4.39 $ & $ 2.91 $  \\
S & $ \mathbf{1.00} $ & $ \mathbf{8} $ & $ \mathbf{318} $ & $ \mathbf{10.95} $ & $ \mathbf{36.64} $ & $ \mathbf{4.41} $ & $ \mathbf{3.34} $ & $ 1.00 $ & $ 24 $ & $ 378 $ & $ 8.52 $ & $ 21.06 $ & $ 2.48 $ & $ 2.47 $  \\
& $ 1.00 $ & $ 8 $ & $ 328 $ & $ 10.77 $ & $ 62.13 $ & $ 14.35 $ & $ 5.76 $ & $ 1.00 $ & $ 28 $ & $ 318 $ & $ 9.32 $ & $ 18.54 $ & $ 2.00 $ & $ 1.99 $  \\
& $ 1.00 $ & $ 8 $ & $ 338 $ & $ 10.50 $ & $ 52.30 $ & $ 9.98 $ & $ 4.98 $ & $ 1.00 $ & $ 32 $ & $ 318 $ & $ 9.28 $ & $ 16.75 $ & $ 2.46 $ & $ 1.80 $  \\
& $ 1.00 $ & $ 8 $ & $ 348 $ & $ 10.30 $ & $ 47.97 $ & $ 7.21 $ & $ 4.65 $ & $ 1.25 $ & $ 16 $ & $ 318 $ & $ 10.19 $ & $ 33.69 $ & $ 10.99 $ & $ 3.30 $  \\
& $ 1.00 $ & $ 8 $ & $ 358 $ & $ 10.17 $ & $ 53.44 $ & $ 19.78 $ & $ 5.25 $ & $ 1.50 $ & $ 16 $ & $ 308 $ & $ 10.13 $ & $ 25.21 $ & $ 2.01 $ & $ 2.48 $  \\
& $ 1.00 $ & $ 8 $ & $ 368 $ & $ 10.08 $ & $ 63.59 $ & $ 3.03 $ & $ 6.30 $ & $ 1.50 $ & $ 16 $ & $ 318 $ & $ 10.30 $ & $ 24.12 $ & $ 2.57 $ & $ 2.34 $  \\
& $ 1.00 $ & $ 8 $ & $ 378 $ & $ 10.07 $ & $ 54.52 $ & $ 3.27 $ & $ 5.41 $ & $ 1.50 $ & $ 16 $ & $ 328 $ & $ 10.14 $ & $ 25.50 $ & $ 3.35 $ & $ 2.51 $  \\
& $ 1.00 $ & $ 12 $ & $ 318 $ & $ 10.53 $ & $ 32.81 $ & $ 3.36 $ & $ 3.11 $ & $ 1.50 $ & $ 16 $ & $ 338 $ & $ 9.45 $ & $ 22.46 $ & $ 7.07 $ & $ 2.37 $  \\
& $ 1.00 $ & $ 16 $ & $ 308 $ & $ 10.18 $ & $ 27.35 $ & $ 2.36 $ & $ 2.68 $ & $ 1.50 $ & $ 16 $ & $ 348 $ & $ 9.56 $ & $ 25.77 $ & $ 4.00 $ & $ 2.69 $  \\
T & $ \mathbf{1.00} $ & $ \mathbf{16} $ & $ \mathbf{318} $ & $ \mathbf{10.15} $ & $ \mathbf{24.94} $ & $ \mathbf{4.33} $ & $ \mathbf{2.45} $ & $ 1.50 $ & $ 16 $ & $ 358 $ & $ 9.44 $ & $ 30.21 $ & $ 3.04 $ & $ 3.20 $  \\
& $ 1.00 $ & $ 16 $ & $ 328 $ & $ 9.99 $ & $ 27.95 $ & $ 4.33 $ & $ 2.79 $ & $ 1.50 $ & $ 16 $ & $ 368 $ & $ 9.23 $ & $ 26.78 $ & $ 4.89 $ & $ 2.90 $  \\
& $ 1.00 $ & $ 16 $ & $ 338 $ & $ 9.85 $ & $ 25.44 $ & $ 2.55 $ & $ 2.58 $ & $ 1.50 $ & $ 16 $ & $ 378 $ & $ 9.19 $ & $ 26.05 $ & $ 2.74 $ & $ 2.83 $  \\
& $ 1.00 $ & $ 16 $ & $ 348 $ & $ 9.65 $ & $ 31.90 $ & $ 4.66 $ & $ 3.30 $ & $ 1.75 $ & $ 16 $ & $ 318 $ & $ 10.37 $ & $ 24.91 $ & $ 2.26 $ & $ 2.40 $  \\
& $ 1.00 $ & $ 16 $ & $ 358 $ & $ 9.58 $ & $ 31.63 $ & $ 5.21 $ & $ 3.30 $ & $ 2.00 $ & $ 16 $ & $ 318 $ & $ 10.89 $ & $ 26.21 $ & $ 4.37 $ & $ 2.40 $  \\

\tableline
\end{tabular}
\label{table_ts_all}
\end{center}
\end{table}

When $\mathrm{Rm}$ is fixed to its standard value (model S, $\mathrm{Rm}=318$), 
 our results show that $\tau_e$ and $T_c$ are mostly sensitive to
 $\mathrm{C}_{\alpha}$, and less affected by variations in $\mathrm{C}_{BL}$. 
 Regarding the former dependency, it can be seen in Fig. \ref{fig_time_scales_CaCs}(\textit{a}) that 
 both $\tau_e$ and $T_c$ decrease with $\mathrm{C}_\alpha$. The decrease in $\tau_e$ is a consequence of
 the stronger destabilizing effect of turbulence. 
 We also note (Fig.~\ref{fig_time_scales_CaCs}(\textit{a}), right) that the ratio $\tau_e/T_c$ 
 decreases with increasing $\mathrm{C}_\alpha$. In the parameter region
 which we explored, $\tau_e$ is thus more sensitive 
 to variations in $\mathrm{C}_\alpha$ than $T_c$.  

 According to Fig.~\ref{fig_time_scales_CaCs}(\textit{b}), the cycle period $T_c$ displays a non-monotonic behavior 
with respect to changes in $\mathrm{C}_{BL}$, which measures the intensity of the 
non-local coupling in the governing Eq.~(\ref{induction_pol}). It is worth mentionning
here that for the lower value of $\mathrm{C}_{BL}$,  
the system undergoes a transition to an $\alpha$-dominated
dynamo, characterized by a longer (and less solar-like) periodicity of about $13$~years. 
 As indicated by 
 Fig.~\ref{fig_time_scales_CaCs}(\textit{b}), 
 the $e$-folding time $\tau_e$ does not vary substantially 
  with $\mathrm{C}_{BL}$ over our narrow interval of investigation (recall section~\ref{section_sensitivity}). Overall, we find that the
    ratio $\tau_e/T_c$ remains approximately constant (equal to $2.5$) over this interval.

Turning our attention to the dependency of $T_c$ and $\tau_e$ on 
$\mathrm{Rm}$, we see (Fig.~\ref{fig_time_scales_Rm}) that the former decreases
 with increasing  $\mathrm{Rm}$. 
The cycle duration scales indeed approximately in inverse proportion to 
$\mathrm{Rm}$, as shown in the left
 panel of Fig.~\ref{fig_time_scales_Rm}. On the other hand, the dependency of $\tau_e$ on $\mathrm{Rm}$ is 
 less clear. There seems to be a mild trend in the cases of low to intermediate 
values of $\mathrm{C}_{\alpha}$ (orange and red points in the 
middle plot of Fig.~\ref{fig_time_scales_Rm}), with $\tau_e$
slightly
increasing with increasing $\mathrm{Rm}$.   This behavior can be 
interpreted as  a regulatory effect of the meridional circulation: 
 as $u_0$ gets larger the meridional circulation 
 tends to make the system more stable against perturbations.
 This is no longer true for a large  $\mathrm{C}_{\alpha}$ (dark red points in the 
 middle plot of  Fig.~\ref{fig_time_scales_Rm}), which indicates
  that $\tau_e$ is then controlled by the $\alpha$-effect. 
  
   It is worth mentionning that some realisations of $\tau_e$ are affected by large uncertainties, 
  mostly in cases with low values of $\mathrm{C}_{\alpha}$ and $\mathrm{C}_{BL}$, and large
  values of $\mathrm{Rm}$. These cases are the less chaotic ones, and the introduction of a 
  perturbation can sometimes lead to a mobilization phase lasting for more than $1,000$~years.


To conclude this analysis, let us stress (as shown in
Fig.~\ref{fig_time_scales_Rm}, right) that
the $\tau_e/T_c$ ratio is mainly concentrated around two values, 2.5 and 5, with 
a larger concentration of points around the former. Using all the available data at our disposal (as summarized in Table \ref{table_ts_all}),
we can finally calculate a weighted average for the ratio  $\tau_e/T_c$, and find 
\begin{eqnarray}
\frac{\tau_e}{T_c} = 2.76 \pm 0.05.
\end{eqnarray}


\section{Summary and discussion} \label{section_discussion}

Our extensive analysis of the $e$-folding time $\tau_e$ for our preferred (in the sense of solar semblance)
standard model S led us to conclude
 that if the control 
 parameters $(\mathrm{Rm},\mathrm{C}_\alpha,\mathrm{C}_{BL})$ 
 are fixed, then  
  $\tau_e$ can be regarded as an intrinsic property of the model, regardless of the source of the error, 
  with a small dependence on its initial magnitude (section~\ref{sec:syst_pert}). 

In view of using that standard model (or a close version) for operational
forecasting, we extended the analysis to a series of models, and investigated  
the sensitivity of $\tau_e$ to $(\mathrm{Rm},\mathrm{C}_\alpha,\mathrm{C}_{BL})$ in detail. 
 Our results  reveal three salient properties  
\begin{enumerate}
\item a decrease of $\tau_e$ with increasing $\mathrm{C}_{\alpha}$.
This reflects the influence of the non-linear nature of the quenched $\alpha$-effect 
on the amplification of errors, 
leading to a more chaotic (and less predictable) dynamo;
\item an apparent independence of  $\tau_e$ on $\mathrm{C}_{BL}$, indicating the secondary role played
by this non-local forcing term on the error growth. However, let us stress that this may be
caused by 
the narrow range of possible $\mathrm{C}_{BL}$ we explored, a consequence of the extreme sensitivity of
the solar semblance of the flux-transport model to this parameter;
\item  a slight tendency for $\tau_e$ to increase with $\mathrm{Rm}$ for those
 models with low to intermediate strength of the $\alpha$-effect,  pointing to a stabilizing  role of the meridional circulation on the system under these conditions. 
\end{enumerate}
In addition, the moderate variability of the ratio of $\tau_e$ to the simulated cycle period 
$T_c$ in our database of simulations (which comprises approximately $50$ members) prompts us to
propose the master value $\overline{\tau}_e = 2.76$~$T_c$ for the class of mean-field models we considered, 
should they be used for operational forecasting (and keeping in mind that
we focussed our analysis on the regular working of those models, not considering extreme
events such as grand minima). 

From a practical point of view, the perturbations artificially inserted 
  into the model in 
section~\ref{section_eg} can be interpreted as uncertainties
in the measurements or in the model itself, which are the causes of errors any data assimilation 
scheme needs to deal with. These uncertainties are ultimately responsible for the limited 
horizon of predictability of the chaotic system we are interested in. If $\varepsilon$ denotes
the relative level of these uncertainties, we derive from Eq.~(\ref{eq:eg}) that the forecast horizon $\tau_f$ is given by
\begin{eqnarray} \label{eq:tauf} 
\tau_f = - \overline{\tau}_e \ln \varepsilon. 
\end{eqnarray}

\begin{figure}[!h]
\epsscale{0.65}
\plotone{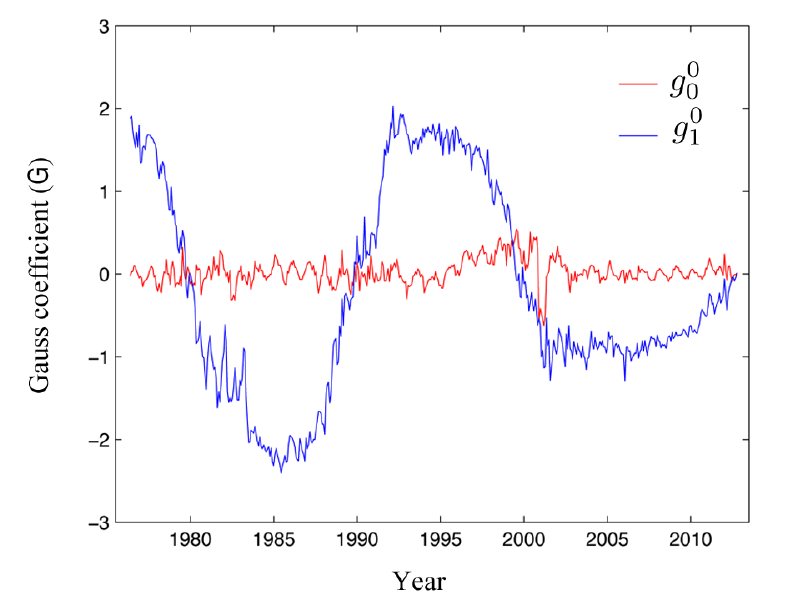}
\caption{Time series of the
monopole $g_0^0$ and axial dipole $g_1^0$ Gauss coefficients derived 
from the magnetic charts of the Wilcox Solar Observatory 
.}
\label{fig_gauss_coef}
\end{figure}

Let us begin by estimating the level of uncertainties on the measurement side. It is likely
that an operational data assimilation scheme will assimilate observations
connected with large-scale maps of $B_r$ at the solar surface, $B_r^S$. Such magnetograms are
contaminated by errors, due to limited resolution, asynchronous sampling and sparse polar measurements. A way to quantify those errors 
is to analyze the spherical harmonic decomposition of $B_r^S$. Theory demands
the monopole term in this expansion ($g_0^0$) to be zero; a non-zero 
$g_0^0$ can consequently be used as a means to quantify the uncertainty $\varepsilon\left(B_r^S\right)$ we are after.  
Figure~\ref{fig_gauss_coef} shows the time series of the monopole and axial
dipole coefficient ($g_1^0$) derived from the database of magnetograms of 
the Wilcox Solar Observatory (WSO)\footnote{http://wso.standford.edu/Harmonic.rad/ghlist.html}. The figure shows that 
$g_1^0$ evolves in phase with the global 
poloidal magnetic field -- it changes sign at the time of maximum activity, 
and is correlated with the polar flux \citep{derosa2012solar}. 
The monopole coefficient $g_0^0$ constantly oscillates around zero. We can 
therefore use the ratio 
of the root-mean-squared (rms) value of $g_0^0$, $\langle g_0^0 \rangle$, to the rms value of $g_1^0$, 
$\langle g_1^0 \rangle$, 
to estimate
$\varepsilon\left(B_r^S\right)$. This yields
\begin{eqnarray}
\varepsilon\left(B_r^S \right) \approx \frac{\langle g_0^0 \rangle}{\langle g_1^0 \rangle} = 
\frac{0.1535\mbox{ G}}{1.2550 \mbox{ G}} \approx 12 \%. 
\end{eqnarray}

On the model side now, one of the most obvious sources of errors lies in the large-scale
kinematic approximation on which our modelling rests. 
In particular, observations
indicate that it may be inappropriate to assume that the large-scale flow driving the dynamo is steady. 
As a consequence, there are errors arising from the 
variability of both the patterns of differential rotation and meridional circulation.
\citet{howe2000dynamic} discovered a persistent pattern of low-amplitude time variation of 
$\Omega$, $\delta \Omega$, of about $6$~nHz, due to solar torsional oscillations. 
Consequently, we get
\begin{eqnarray}
\varepsilon(\Omega) = \frac{\delta \Omega}{\Omega_{eq}} \approx 1 \%. 
\end{eqnarray}
This small figure must be contrasted with the one
owing to those 
uncertainties impacting ${\mathbf u}_p$. 
The long-term variability of the meridional circulation 
$\delta {\mathbf u}_p$ has an amplitude $\delta u_0$  close to $5$~m/s \citep{hernandez2006meridional}, which yields
\begin{eqnarray}
\varepsilon(\mathbf{u}_p) = \frac{\delta u_0}{u_0} \approx 33 \%,  
\end{eqnarray}
if computed based on the mean value of the surface 
meridional flow at mid-latitudes, $u_0~\approx~15$~m/s. 
Injecting $\varepsilon(\mathbf{u}_p)$ in Eq.~(\ref{eq:tauf}) yields 
\begin{eqnarray}
\tau_f \approx 3 \ T_c. 
\end{eqnarray} 
In addition to these fluctuations in amplitude, 
there exists considerable uncertainties on the large-scale structure of the meridional 
circulation itself. 
The depth at which the equatorward return flow occurs \citep{hathaway2011sun} and 
the possible multi-cell pattern of meridional flow \citep{zhao2013detection} are 
two examples illustrating the current lack of 
robust observational constraints on ${\mathbf u}_p$. These can not be readily
 incorporated in the current analysis, for they would require
 different families of simulations to be integrated, and their region 
 of solar semblance be identified in parameter space (in the same way 
 we mapped it for the ensemble of single-cell, tachocline equatorward return flow
 simulations 
 considered here). 




Further uncertainties affect the turbulent diffusivity, $\eta({\mathbf r})$. 
As explained in Sec.~\ref{section_model}, we resort in this study to 
spherically symmetric $\eta_p(r)$ and $\eta_t(r)$, of relatively 
low values. Both reach an amplitude of $3\times10^{11}$ cm/s$^{2}$
at $r=R_{\sun}$ (recall Fig.~\ref{fig_flow}({\it d})). This value has to be constrasted with 
the value of $10^{12}$ cm/s$^{2}$ used by \cite{choudhuri2007predicting}
for their prediction of cycle~24, using a so-called diffusion-dominated
 flux-transport model. More recently, \cite{miesch2012amplitude}
put forward theoretical and observational arguments in favor
of the same figure, $10^{12}$ cm/s$^{2}$, as a lower bound 
of this turbulent transport coefficient. The exact 
nature of $\eta({\mathbf r})$ in the solar interior 
remains strongly debated, and no consensus has been reached. 

In this study, we opted for an advection-dominated forward model, on
the account of its first order dynamical semblance with the solar dynamo. 
If one were to choose instead a diffusion-dominated model for data assimilation 
purposes, one would have to carry out a sensitivity analysis
 similar to the one pursued here, in order to compute
  the $e$-folding time and estimate the forecast horizon for that
different family of models. In this respect, note that 
 \citet{karak2012turbulent} resorted to a simpler, correlation-based, analysis 
  in order to study 
the ``memory" of such a diffusion-dominated model (which includes
in their case turbulent pumping and a stochastic component to the poloidal 
source term). Their analysis demonstrates that turbulent diffusion
shortens the memory of the system to less than one cycle. This
preliminary work should be complemented by the proper derivation
of the $e$-folding time characterizing that class of models, along
the methodological lines presented in this paper. 


Regardless of the model ultimately chosen, one should keep in mind that data assimilation
remains in any case the only sensible way of testing the compatibility
of a given physical model of the solar dynamo with observations 
of its dynamical activity. By enabling on-the-fly 
parameter adjustments (in addition to state estimation), data
assimilation offers in principle the possibility of correcting
 the radial profiles of diffusion coefficients (and those of poloidal source terms). 
 Even if the advection-dominated model we studied has an optimistic theoretical predictability limit of
 three solar cycles, we must bear in mind that any data assimilation scheme aiming to forecast solar
 activity will be unperfect, and its effective forecast horizon will consequently decrease.
Taking this into account, one can hope, though,
that if such an advection-dominated model were to be chosen for operational forecasting, 
  its practical
  limit of predictability could reach (and perhaps exceed) one solar cycle.

\clearpage

\section{Acknowledgments}

The authors would like to thank the referee for his/her helpful and constructive review, 
and Allan Sacha Brun, Emmanuel Dormy and Martin Schrinner for enlightening 
discussions. Sabrina Sanchez would like to thank as well 
Oscar Matsuura and Katia Pinheiro for the fruitful 
contributions to the beginning of this project, the Observat\'orio Nacional of Brazil for the initial 
support,  and the Space Physics and Aeronomy group of the American Geophysical Union for the student
grant award at the 2012 AGU Meeting.
Numerical calculations were performed 
on IPGP's S-CAPAD computing facility. This is IPGP contribution 3459.

\clearpage


\appendix
\counterwithin{figure}{section}
\counterwithin{table}{section}
\section{Parody Code - Mean Field Benchmarking}
\label{appendix_1}

The Parody code used in this work was originally proposed for full 3D MHD dynamo simulations (ACD code, 
benchmarked in \citet{christensen2001numerical}, see \cite{dormy1998mhd} and \cite{aubert2008magnetic}).
In order to perform an analysis of the predictability of standard mean-field solar dynamos,
it was necessary to ensure the compatibility of the model with 
the ones used within 
the solar dynamo community. For such reason, we modified and compared outputs from our 3D MHD
code with a mean-field solar dynamo benchmark.

The full spherical harmonic expansion of the code writes
\begin{eqnarray} \label{B_sh_full}
({\mathcal P},{\mathcal T})(r,\theta,\varphi,t)
= \sum_{n = 1}^{N} \sum_{m = 1}^{M} ({\mathcal P}_n^m, {\mathcal T}_n^m)(r,t) \; Y_{n}^m (\theta,\varphi),
\end{eqnarray}
truncated at spherical harmonic degree and order $N$ and $M$, respectivelly. As most mean-field models assume axisymmetry, we set $M=0$ throughout. 

The original inner boundary conditions of Parody considered the
inner core as an insulating or electrically conducting medium of finite conductivity 
\citep{christensen2001numerical}; in contrast, in 
the solar context, the radiative zone is modelled as a perfect conductor. 
 This last condition requires to impose 
\begin{eqnarray} 
{\mathcal P}&=&0, \mbox{ and} \\
\partial(r \,{\mathcal T})/\partial r &=&0 \mbox{ at the inner boundary.} 
\end{eqnarray} 
 Further 
modifications of the code included the incorporation of the $\alpha$ and BL source terms in the poloidal induction
Equation~\eqref{induction_pol}, and the prescription of the flow fields, $\Omega$ and $\mathbf{u}_p$, and depth-dependent turbulent diffusivities $\eta(r)$.

\begin{figure}[h!]
\renewcommand{\figurename}{Appendix}
\epsscale{.80}
\plotone{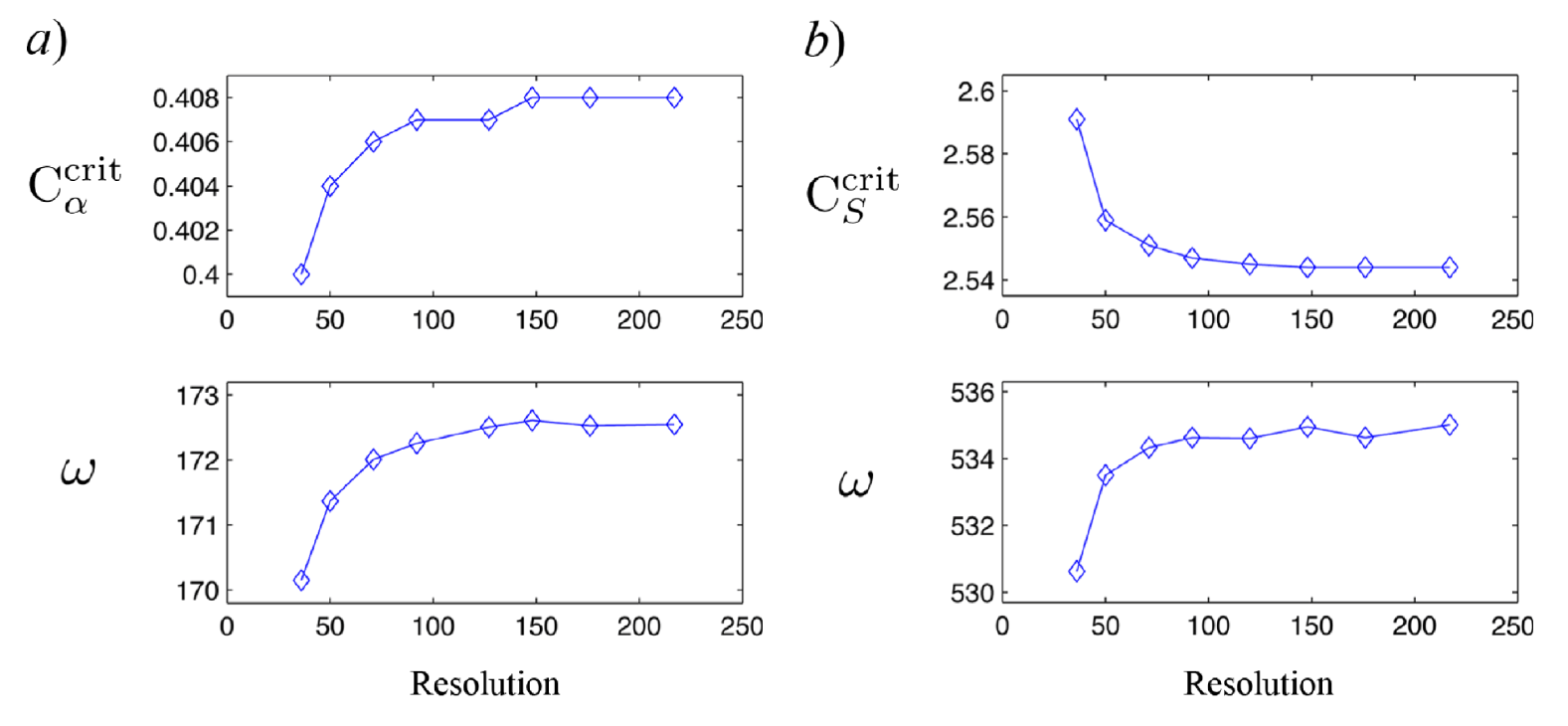}
\caption{Convergence tests: critical dynamo numbers $\mathrm{C}^{\mathrm{crit}}_\alpha$ and $\mathrm{C}^{\mathrm{crit}}_S$ and solar cycle periodicity $\omega$ 
for \textit{a}) case B and \textit{b}) case C, as defined by \citet{jouve2008solar}. The resolution is defined by 
$\sqrt{N_r\, N}$, where $N_r$ is the number of radial levels and $N$ is the truncation of the spherical
harmonic expansion.}
\label{fig_bench_conv}
\end{figure}

\begin{figure}[t!]
\renewcommand{\figurename}{Appendix}
\epsscale{.80}
\plotone{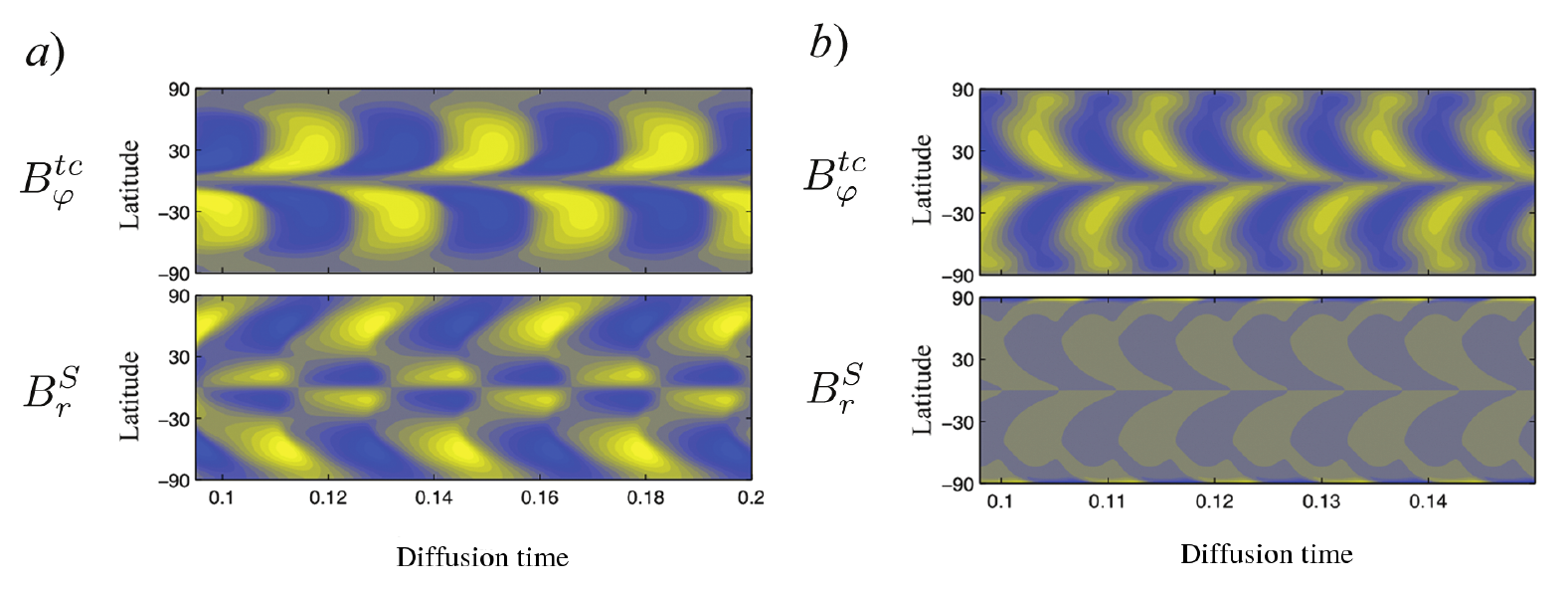}
\caption{Butterfly diagrams summarizing two different benchmark cases from \citet{jouve2008solar}: \textit{a}) $\alpha\Omega$ dynamo from the supercritical case SB and \textit{b}) a Babcock-Leighton 
dynamo from the supercritical case SC. For each case, the upper panel
displays the toroidal field at the tachocline and the lower one the radial field at the surface.}
\label{fig_bench_btfly}
\end{figure}

The resulting code was tested against published reference solutions of the mean-field 
 community benchmark effort described by \cite{jouve2008solar}. The
benchmarking consists of computing the critical dynamo numbers $\mathrm{C^{crit}}$, and solar activity 
cycle frequency $\omega$, for three case studies. The three cases include two
$\alpha\Omega$ mean-field dynamos (cases A and B, differing only by the prescribed $\eta(r)$) and a 
Babcock-Leighton dynamo (case C). Table \ref{table_bench} displays the 
values obtained from our code and the \citet{jouve2008solar} benchmark ones for each case, while  
convergence tests of the critical numbers of cases B and C are shown in~Figure~\ref{fig_bench_conv}. In addition, 
butterfly diagrams for the supercritical cases SB and SC (the supercritical cases include $\alpha$ and $S_{BL}$ quenching) are displayed in Figure~\ref{fig_bench_btfly}.

\begin{table}[!ht]
\begin{center}
\scriptsize
\caption{Comparison of the critical dynamo numbers $\mathrm{C}_{\alpha,S}^{\mathrm{crit}}$ and 
frequency of the solar cycle $\omega$
in the benchmark cases A, B and C from \citet{jouve2008solar}. The spacial
and temporal resolutions are given in terms of radial points and harmonic degree 
($N_r \times N$) and time-step $\Delta t$.}
\begin{tabular} {ccccccc}
\tableline \tableline

 & \multicolumn{4}{c}{Results} & \multicolumn{2}{c}{Reference} \\
 Case & Resolution & $\Delta t$ & $\mathrm{C}_{\alpha,s}^{\mathrm{crit}}$ & $\omega$ & $\mathrm{C}_{\alpha,s}^{\mathrm{crit}}$ & $\omega$  \\
\tableline

A    & 71 $\times$ 71   & $5\times10^{-5}$ & 0.385 & 158.00 & 0.387 $\pm$ 0.002 & 158.1 $\pm$ 1.472 \\
B    & 71 $\times$ 71   & $5\times10^{-5}$ & 0.406 & 172.01 & 0.408 $\pm$ 0.003 & 172.0 $\pm$ 0.632 \\
C    & 120 $\times$ 120 & $ 1\times10^{-6}$ & 2.545 & 534.6 & 2.489 $\pm$ 0.075 & 536.6 $\pm$ 8.295 \\

\tableline
\end{tabular}
\label{table_bench}
\end{center}
\end{table}

Note that in the present study, and compared with the 
benchmark cases, we use slightly different inner boundary conditions, namely 
\begin{eqnarray}
{\mathcal P}&=&0, \mbox{ and} \\
{\mathcal T}&=&0 \mbox{ at the inner boundary}, 
\end{eqnarray}
 as is common 
in mean-field solar dynamo simulations \citep[e.g.][]{dikpati1999babcock}.
Inspection of results obtained with both types of inner boundary conditions shows that they are virtually 
 the same, in agreement with \citet{chatterjee2004full}. An interpretation of this is that the
low diffusivity of the radiative zone and the absence of a deeply penetrating meridional flow 
 inhibit the penetration of the strong tachocline magnetic field to the deepermost layers.


\clearpage

\bibliography{biblio_solar}

\newcommand{\noop}[1]{}
\begin{thebibliography}{}

\bibitem[Aubert et~al., 2008]{aubert2008magnetic}
Aubert, J., Aurnou, J., and Wicht, J. (2008).
\newblock The magnetic structure of convection-driven numerical dynamos.
\newblock {\em Geophysical Journal International}, 172(3):945--956.

\bibitem[Aubert and Fournier, 2011]{aubert2011npg}
Aubert, J. and Fournier, A. ({2011}).
\newblock {Inferring internal properties of {E}arth's core dynamics and their
  evolution from surface observations and a numerical geodynamo model}.
\newblock {\em {Nonlinear Processes In Geophysics}}, {18}({5}):{657--674}.

\bibitem[Babcock, 1961]{babcock1961topology}
Babcock, H.~W. (1961).
\newblock The topology of the {S}un's magnetic field and the 22-year cycle.
\newblock {\em The Astrophysical Journal}, 133:572.

\bibitem[Baker, 2000]{baker2000occurrence}
Baker, D.~N. (2000).
\newblock The occurrence of operational anomalies in spacecraft and their
  relationship to space weather.
\newblock {\em Plasma Science, IEEE Transactions on}, 28(6):2007--2016.

\bibitem[Bonanno et~al., 2002]{bonanno2002parity}
Bonanno, A., Elstner, D., R{\"u}diger, G., and Belvedere, G. (2002).
\newblock Parity properties of an advection-dominated solar dynamo.
\newblock {\em Astronomy and Astrophysics}, 390(2):673--680.

\bibitem[Brasseur, 2006]{brasseur06}
Brasseur, P. (2006).
\newblock Ocean data assimilation using sequential methods based on the
  {K}alman filter.
\newblock In Chassignet, E. and Verron, J., editors, {\em Ocean Weather
  Forecasting: An Integrated View of Oceanography}, pages 271--316. Springer.

\bibitem[Brun et~al., 2004]{brun2004global}
Brun, A.~S., ~, M.~S., and Toomre, J. (2004).
\newblock Global-scale turbulent convection and magnetic dynamo action in the
  solar envelope.
\newblock {\em The Astrophysical Journal}, 614(2):1073.

\bibitem[Brun, 2007]{brun2007towards}
Brun, A.~S. (2007).
\newblock Towards using modern data assimilation and weather forecasting
  methods in solar physics.
\newblock {\em Astronomische Nachrichten}, 328(3-4):329--338.

\bibitem[Bushby and Tobias, 2007]{bushby2007predicting}
Bushby, P.~J. and Tobias, S.~M. (2007).
\newblock On predicting the solar cycle using mean-field models.
\newblock {\em The Astrophysical Journal}, 661(2):1289.

\bibitem[Charbonneau, 2005]{charbonneau2010dynamo}
Charbonneau, P. (2005).
\newblock Dynamo models of the solar cycle.
\newblock {\em Living Reviews in Solar Physics}, 2(2).

\bibitem[Charbonneau and Dikpati, 2000]{charbonneau2000stochastic}
Charbonneau, P. and Dikpati, M. (2000).
\newblock Stochastic fluctuations in a {B}abcock-{L}eighton model of the solar
  cycle.
\newblock {\em The Astrophysical Journal}, 543(2):1027.

\bibitem[Charbonneau and Smolarkiewicz, 2013]{charbonneau2013modeling}
Charbonneau, P. and Smolarkiewicz, P.~K. (2013).
\newblock Modeling the {S}olar {D}ynamo.
\newblock {\em Science}, 340(6128):42--43.

\bibitem[Charbonneau et~al., 2005]{charbonneau2005fluctuations}
Charbonneau, P., St-Jean, C., and Zacharias, P. (2005).
\newblock Fluctuations in {B}abcock-{L}eighton dynamos. {I}. {P}eriod doubling
  and transition to chaos.
\newblock {\em The Astrophysical Journal}, 619(1):613.

\bibitem[Chatterjee et~al., 2004]{chatterjee2004full}
Chatterjee, P., Nandy, D., and Choudhuri, A.~R. (2004).
\newblock Full-sphere simulations of a circulation-dominated solar dynamo:
  {E}xploring the parity issue.
\newblock {\em Astronomy and Astrophysics}, 427(3):1019--1030.

\bibitem[Choudhuri et~al., 2007]{choudhuri2007predicting}
Choudhuri, A.~R., Chatterjee, P., and Jiang, J. (2007).
\newblock Predicting solar cycle 24 with a solar dynamo model.
\newblock {\em Physical review letters}, 98(13):131103.

\bibitem[Christensen et~al., 2001]{christensen2001numerical}
Christensen, U.~R., Aubert, J., Cardin, P., Dormy, E., Gibbons, S., Glatzmaier,
  G.~A., Grote, E., Honkura, Y., Jones, C., Kono, M., et~al. (2001).
\newblock A numerical dynamo benchmark.
\newblock {\em Physics of the Earth and Planetary Interiors}, 128(1):25--34.

\bibitem[DeRosa et~al., 2012]{derosa2012solar}
DeRosa, M.~L., Brun, A.~S., and Hoeksema, J.~T. (2012).
\newblock Solar magnetic field reversals and the role of dynamo families.
\newblock {\em The Astrophysical Journal}, 757(1):96.

\bibitem[Dikpati and Anderson, 2012]{dikpati2012evaluating}
Dikpati, M. and Anderson, J.~L. (2012).
\newblock Evaluating potential for data assimilation in a flux-transport dynamo
  model by assessing sensitivity and response to meridional flow variation.
\newblock {\em The Astrophysical Journal}, 756(1):20.

\bibitem[Dikpati and Charbonneau, 1999]{dikpati1999babcock}
Dikpati, M. and Charbonneau, P. (1999).
\newblock A {B}abcock-{L}eighton flux transport dynamo with solar-like
  differential rotation.
\newblock {\em The Astrophysical Journal}, 518(1):508.

\bibitem[Dikpati et~al., 2006]{dikpati2006predicting}
Dikpati, M., De~Toma, G., and Gilman, P.~A. (2006).
\newblock Predicting the strength of solar cycle 24 using a flux-transport
  dynamo-based tool.
\newblock {\em Geophysical research letters}, 33(5):L05102.

\bibitem[Dikpati and Gilman, 2001]{dikpati2001flux}
Dikpati, M. and Gilman, P.~A. (2001).
\newblock Flux-transport dynamos with $\alpha$-effect from global instability
  of tachocline differential rotation: a solution for magnetic parity selection
  in the {S}un.
\newblock {\em The Astrophysical Journal}, 559(1):428.

\bibitem[Dormy et~al., 1998]{dormy1998mhd}
Dormy, E., Cardin, P., and Jault, D. (1998).
\newblock {MHD} flow in a slightly differentially rotating spherical shell,
  with conducting inner core, in a dipolar magnetic field.
\newblock {\em Earth and Planetary Science Letters}, 160(1):15--30.

\bibitem[D'{S}ilva and Choudhuri, 1993]{dsilva1993theoretical}
D'{S}ilva, S. and Choudhuri, A.~R. (1993).
\newblock A theoretical model for tilts of bipolar magnetic regions.
\newblock {\em Astronomy and Astrophysics}, 272:621.

\bibitem[Elbern et~al., 2010]{elbern2010inverse}
Elbern, H., Strunk, A., and Nieradzik, L. (2010).
\newblock Inverse modelling and combined state-source estimation for chemical
  weather.
\newblock In {\em Data Assimilation}, pages 491--513. Springer.

\bibitem[Fan, 2009]{fan2009magnetic}
Fan, Y. (2009).
\newblock Magnetic fields in the solar convection zone.
\newblock {\em Living Reviews in Solar Physics}, 6(4).

\bibitem[Fournier et~al., 2007]{fournier2007case}
Fournier, A., Eymin, C., and Alboussi{\`e}re, T. (2007).
\newblock A case for variational geomagnetic data assimilation: insights from a
  one-dimensional, nonlinear, and sparsely observed {MHD} system.
\newblock {\em Nonlinear processes in Geophysics}, 14:163--180.

\bibitem[Fournier et~al., 2010]{fournier2010introduction}
Fournier, A., Hulot, G., Jault, D., Kuang, W., Tangborn, A., Gillet, N., Canet,
  E., Aubert, J., and Lhuillier, F. (2010).
\newblock An introduction to data assimilation and predictability in
  geomagnetism.
\newblock {\em Space science reviews}, 155(1-4):247--291.

\bibitem[Fournier et~al., 2013]{fournier2013ensemble}
Fournier, A., Nerger, L., and Aubert, J. (2013).
\newblock An ensemble {K}alman filter for the time-dependent analysis of the
  geomagnetic field.
\newblock {\em Geochemistry, Geophysics, Geosystems}, 14.
\newblock doi:10.1002/ggge.20252.

\bibitem[Haigh, 2003]{haigh2003effects}
Haigh, J.~D. (2003).
\newblock The effects of solar variability on the {E}arth's climate.
\newblock {\em Philosophical Transactions of the Royal Society of London.
  Series A: Mathematical, Physical and Engineering Sciences},
  361(1802):95--111.

\bibitem[Hathaway, 1996]{hathaway1996doppler}
Hathaway, D.~H. (1996).
\newblock Doppler measurements of the {S}un's meridional flow.
\newblock {\em The Astrophysical Journal}, 460:1027.

\bibitem[Hathaway, 2009]{hathaway2009solar}
Hathaway, D.~H. (2009).
\newblock Solar cycle forecasting.
\newblock {\em Space science reviews}, 144(1-4):401--412.

\bibitem[Hathaway, 2010]{hathaway2010solar}
Hathaway, D.~H. (2010).
\newblock The solar cycle.
\newblock {\em Living Reviews in Solar Physics}, 7(1).

\bibitem[Hathaway, 2011]{hathaway2011sun}
Hathaway, D.~H. (2011).
\newblock The sun's shallow meridional circulation.
\newblock {\em arXiv preprint arXiv:1103.1561}.

\bibitem[Hern{\'a}ndez et~al., 2006]{hernandez2006meridional}
Hern{\'a}ndez, I.~G., Komm, R., Hill, F., Howe, R., Corbard, T., and Haber,
  D.~A. (2006).
\newblock Meridional circulation variability from large-aperture ring-diagram
  analysis of global oscillation network group and {M}ichelson {D}oppler imager
  data.
\newblock {\em The Astrophysical Journal}, 638(1):576.

\bibitem[Houser et~al., 2010]{houser2010land}
Houser, P.~R., De~Lannoy, G.~J., and Walker, J.~P. (2010).
\newblock Land surface data assimilation.
\newblock In {\em Data Assimilation}, pages 549--597. Springer.

\bibitem[Howe et~al., 2000]{howe2000dynamic}
Howe, R., Christensen-Dalsgaard, J., Hill, F., Komm, R.~W., Larsen, R.~M.,
  Schou, J., Thompson, M.~J., and Toomre, J. (2000).
\newblock Dynamic variations at the base of the solar convection zone.
\newblock {\em Science}, 287(5462):2456--2460.

\bibitem[Hulot et~al., 2010]{hulot2010earth}
Hulot, G., Lhuillier, F., and Aubert, J. (2010).
\newblock Earth's dynamo limit of predictability.
\newblock {\em Geophysical Research Letters}, 37(6).

\bibitem[Jouve et~al., 2008]{jouve2008solar}
Jouve, L., Brun, A.~S., Arlt, R., Brandenburg, A., Dikpati, M., Bonanno, A.,
  K{\"a}pyl{\"a}, P., Moss, D., Rempel, M., Gilman, P., et~al. (2008).
\newblock A solar mean field dynamo benchmark.
\newblock {\em Astronomy and Astrophysics}, 483(3):949--960.

\bibitem[Jouve et~al., 2011]{jouve2011assimilating}
Jouve, L., Brun, A.~S., and Talagrand, O. (2011).
\newblock Assimilating data into an {$\alpha\Omega$} dynamo model of the {S}un:
  A variational approach.
\newblock {\em The Astrophysical Journal}, 735(1):31.

\bibitem[Jouve et~al., 2010]{jouve2010buoyancy}
Jouve, L., Proctor, M. R.~E., and Lesur, G. (2010).
\newblock Buoyancy-induced time delays in {B}abcock-{L}eighton flux-transport
  dynamo models.
\newblock {\em Astronomy and Astrophysics}, 519:13.

\bibitem[Kalnay, 2003]{kalnay2003atmospheric}
Kalnay, E. (2003).
\newblock {\em Atmospheric modeling, data assimilation and predictability}.
\newblock Cambridge university press, Cambridge, UK.

\bibitem[Karak and Nandy, 2012]{karak2012turbulent}
Karak, B.~B. and Nandy, D. (2012).
\newblock Turbulent pumping of magnetic flux reduces solar cycle memory and
  thus impacts predictability of the sun's activity.
\newblock {\em The Astrophysical Journal Letters}, 761(1):L13.

\bibitem[Kitiashvili and Kosovichev, 2008]{kitiashvili2008application}
Kitiashvili, I. and Kosovichev, A.~G. (2008).
\newblock Application of data assimilation method for predicting solar cycles.
\newblock {\em The Astrophysical Journal Letters}, 688(1):L49.

\bibitem[Leighton, 1969]{leighton1969magneto}
Leighton, R.~B. (1969).
\newblock A magneto-kinematic model of the solar cycle.
\newblock {\em The Astrophysical Journal}, 156:1.

\bibitem[Lhuillier et~al., 2011]{lhuillier2011earth}
Lhuillier, F., Aubert, J., and Hulot, G. (2011).
\newblock Earth's dynamo limit of predictability controlled by magnetic
  dissipation.
\newblock {\em Geophysical Journal International}, 186(2):492--508.

\bibitem[Lorenz, 1963]{lorenz1963deterministic}
Lorenz, E.~N. (1963).
\newblock Deterministic nonperiodic flow.
\newblock {\em Journal of the atmospheric sciences}, 20(2):130--141.

\bibitem[Lorenz, 1965]{lorenz1965study}
Lorenz, E.~N. (1965).
\newblock A study of the predictability of a 28-variable atmospheric model.
\newblock {\em Tellus}, 17(3):321--333.

\bibitem[Miesch et~al., 2012]{miesch2012amplitude}
Miesch, M.~S., Featherstone, N.~A., Rempel, M., and Trampedach, R. (2012).
\newblock On the amplitude of convective velocities in the deep solar interior.
\newblock {\em The Astrophysical Journal}, 757(2):128.

\bibitem[Moffatt, 1978]{moffatt1978field}
Moffatt, H.~K. (1978).
\newblock {\em Field Generation in Electrically Conducting Fluids}.
\newblock Cambridge University Press, Cambridge, UK.

\bibitem[Ossendrijver, 2003]{ossendrijver2003solar}
Ossendrijver, M. (2003).
\newblock The solar dynamo.
\newblock {\em The Astronomy and Astrophysics Review}, 11(4):287--367.

\bibitem[Parker, 1955]{parker1955hydromagnetic}
Parker, E.~N. (1955).
\newblock Hydromagnetic dynamo models.
\newblock {\em The Astrophysical Journal}, 122:293.

\bibitem[Priest, 1982]{priest1982solar}
Priest, E.~R. (1982).
\newblock {\em Solar magnetohydrodynamics}.
\newblock Reidel, Boston, MA, USA.

\bibitem[Pulkkinen, 2007]{pulkkinen2007space}
Pulkkinen, T. (2007).
\newblock Space weather: Terrestrial perspective.
\newblock {\em Living Reviews in Solar Physics}, 4(1).

\bibitem[Sanchez et~al., ress]{sanchezrevision}
Sanchez, S.~M., Fournier, A., Pinheiro, K. J.~R., and Aubert, J. (\noop{2014}in
  press).
\newblock A mean-field {B}abcock-{L}eighton solar dynamo model with long-term
  variability.
\newblock {\em Anais da Academia Brasileira de Ci\^{e}ncias}.

\bibitem[Talagrand, 1997]{talagrand1997assimilation}
Talagrand, O. (1997).
\newblock Assimilation of observations, an introduction.
\newblock {\em Journal Meteorological Society of Japan}, 75(2):81--99.

\bibitem[Tomczyk et~al., 1995]{tomczyk1995measurement}
Tomczyk, S., Schou, J., and Thompson, M.~J. (1995).
\newblock Measurement of the rotation rate in the deep solar interior.
\newblock {\em The Astrophysical Journal Letters}, 448(1):L57.

\bibitem[Vallis, 2006]{vallis2006atmospheric}
Vallis, G.~K. (2006).
\newblock {\em Atmospheric and oceanic fluid dynamics: Fundamentals and
  large-scale circulation}.
\newblock Cambridge University Press.

\bibitem[Wang and Sheeley, 2009]{wang2009understanding}
Wang, Y.~M. and Sheeley, N.~R. (2009).
\newblock Understanding the geomagnetic precursor of the solar cycle.
\newblock {\em The Astrophysical Journal Letters}, 694(1):L11.

\bibitem[Zhao et~al., 2013]{zhao2013detection}
Zhao, J., Bogart, R., Kosovichev, A., Duvall~Jr, T., and Hartlep, T. (2013).
\newblock Detection of equatorward meridional flow and evidence of double-cell
  meridional circulation inside the sun.
\newblock {\em The Astrophysical Journal Letters}, 774(2):L29.

\end{thebibliography}

\bibliographystyle{apalike}





\end{document}